\documentclass[aps,prd,twocolumn,superscriptaddress,preprintnumbers,floatfix,nofootinbib,notitlepage,showkeys]{revtex4-1}

\usepackage[utf8]{inputenc}
\usepackage{cancel}
\usepackage{float}
\usepackage{graphicx}
\usepackage{hyperref}
\usepackage{latexsym}
\usepackage{amsmath}
\usepackage{amssymb}
\usepackage{bbm}

\usepackage{ulem}
\usepackage{pdfsync}
\usepackage{epsfig}
\usepackage{epstopdf}
\usepackage{subfigure}
\usepackage{color}
\usepackage{comment}
\usepackage{slashed}

\def\beq{\begin{equation}}
\def\eeq{\end{equation}}
\def\baq{\begin{eqnarray}}
\def\eaq{\end{eqnarray}}

\providecommand{\f}[2]{\frac{{#1}}{{#2}}}
\newcommand{\ee}[1]{\begin{equation}#1\end{equation}}

\newcommand{\bea}{\begin{eqnarray}} 
\newcommand{\eea}{\end{eqnarray}}

\newcommand{\bmp}{\noindent\begin{minipage}{16cm}}
\newcommand{\emp}{\end{minipage}\vskip 7mm} 
\def\lsim{\mathrel{\raise.3ex\hbox{$<$\kern-.75em\lower1ex\hbox{$\sim$}}}}
\def\gsim{\mathrel{\raise.3ex\hbox{$>$\kern-.75em\lower1ex\hbox{$\sim$}}}}

\newcommand{\intron}[1]{}

\def\MP{M_{\rm P}}

\begin{document}
\preprint{IMPERIAL/TP/2018/TM/06}
\title{Spectator Dark Matter}

\author{Tommi Markkanen}
\email{t.markkanen@imperial.ac.uk}
\email{tommi.markkanen@kbfi.ee}
\affiliation{Department of Physics, Imperial College London,\\London, SW7 2AZ, United Kingdom}
\affiliation{Laboratory of High Energy and Computational Physics, National Institute of Chemical Physics and
Biophysics, R\"avala pst. 10, Tallinn, 10143, Estonia}
\author{Arttu Rajantie}
\email{a.rajantie@imperial.ac.uk}
\affiliation{Department of Physics, Imperial College London,\\Blackett Laboratory, London, SW7 2AZ, United Kingdom}
\author{Tommi Tenkanen}
\email{ttenkan1@jhu.edu}
\affiliation{Department of Physics and Astronomy, Johns Hopkins University, \\
Baltimore, MD 21218, United States of America}
\affiliation{Astronomy Unit, Queen Mary University of London, \\
Mile End Road, London, E1 4NS, United Kingdom}

\begin{abstract}
The observed dark matter abundance in the Universe can be fully accounted for by a minimally coupled spectator scalar field that was light during inflation and has sufficiently strong self-coupling. In this scenario, dark matter was produced during inflation by amplification of quantum fluctuations of the spectator field. The self-interaction of the field suppresses its fluctuations on large scales, and therefore avoids isocurvature constraints. The scenario does not require any fine-tuning of parameters. In the simplest case of a single real scalar field, the mass of the dark matter particle would be in the range $1~{\rm GeV}\lesssim m\lesssim 10^8~{\rm GeV}$, depending on the scale of inflation, and the lower bound for the quartic self-coupling is $\lambda\gtrsim 0.45$.

\end{abstract}

%
\maketitle

%
\section{Introduction}

The existence of a significant dark matter (DM) component in the Universe seems indisputable \cite{Bertone:2016nfn,Aghanim:2018eyx}. However, due to the increasingly tight constraints on conventional particle DM models \cite{Arcadi:2017kky}, fresh ideas are needed to explain the properties and the observed abundance of DM, as well as its formation mechanism in the early Universe. Instead of undergoing usual thermal freeze-out \cite{Kolb:1990vq} or, alternatively, non-thermal freeze-in \cite{McDonald:2001vt,Hall:2009bx,Bernal:2017kxu}, dark matter abundance may have been initiated purely gravitationally either during or after cosmic inflation. This idea dates back to 1980s (see \cite{Ford:1986sy} and e.g. \cite{ Kolb:1998ki,Chung:1998zb,Peebles:1999fz}) but has recently gained increasing attention, see e.g. \cite{Enqvist:2014zqa,Graham:2015rva,Nurmi:2015ema,Garny:2015sjg,Markkanen:2015xuw,Kainulainen:2016vzv,Bertolami:2016ywc,Heikinheimo:2016yds,Cosme:2017cxk,Enqvist:2017kzh,Cosme:2018nly,Graham:2018jyp,Alonso-Alvarez:2018tus,Ema:2018ucl,Fairbairn:2018bsw}.

In this paper, we will focus on the scenario in which dark matter is produced by amplification of the vacuum fluctuations of a scalar field $\chi$ during inflation. We assume that $\chi$ is spectator field, which means that it is light relative to the Hubble rate during inflation, its energy density is subdominant, and that its couplings with the Standard Model degrees of freedom are negligible.
This scenario has been discussed before in Refs.~\cite{Peebles:1999fz,Enqvist:2014zqa,Nurmi:2015ema,Kainulainen:2016vzv,Alonso-Alvarez:2018tus}.

In contrast to DM creation during the reheating epoch \cite{Markkanen:2015xuw,Fairbairn:2018bsw}, if the field(s) responsible for DM production were amplified during inflation, perturbations in the resulting dark matter energy density may not coincide with those in baryonic matter. Observations of the Cosmic Microwave Background radiation (CMB) by the Planck satellite have recently put stringent constraints on the amount of such isocurvature density perturbations at large scales \cite{Akrami:2018odb}, and any dark matter model dealing with inflation has to satisfy these constraints. Indeed, an example of this is the axion dark matter model, where the isocurvature constraints suggest a specific connection between the energy scale of inflation and the axion decay constant \cite{Marsh:2015xka} (see also \cite{Graham:2018jyp,Guth:2018hsa}). In the current scenario, it has been shown previously that a free minimally coupled scalar field would violate the isocurvature constraints~\cite{Alonso-Alvarez:2018tus} (see also \cite{Bertolami:2016ywc}).

In this paper we will study the production of self-interacting DM using the stochastic approach developed in \cite{Starobinsky:1994bd} (see also \cite{Peebles:1999fz,Enqvist:2012xn,Kunimitsu:2012xx,Hardwick:2017fjo}). A similar scenario was originally studied in \cite{Peebles:1999fz} but in this paper we will improve the analysis in several different ways. First and foremost, we refine the analysis of isocurvature perturbations, showing that the recent Planck data is not problematic for the success of the scenario. We will also discuss the effect of DM self-interactions on the evolution of DM number density, as well as the current observational constraints on DM self-interactions that can be inferred from collisions between galaxy clusters. As we will show, the DM production mechanism we will discuss in this paper is sufficiently strong to yield the measured DM abundance for a wide range of masses extending down to sub-GeV ranges and that all observational bounds considered in this paper may be avoided with no fine-tuning of parameters. We will also discuss different ways to test the scenario.

The paper is organised as follows: in Section \ref{sec:CDM}, we show how cold dark matter forms from an inflationary condensate. In Section \ref{sec:isocurvature}, we discuss the isocurvature perturbations inherent to the scenario, and then present the results and observational constraints in Section \ref{results}. Finally, in Section \ref{conclusions}, we conclude with an outlook.

\section{Cold dark matter from an inflationary condensate}
\label{sec:CDM}

\subsection{Production during inflation}
Our set-up will be one with a sufficiently decoupled scalar $\chi$ with the potential 
\ee{V(\chi)=\frac{1}{2}m^2 \chi^2 +\frac{\lambda}{4}\chi^4\,,\label{eq:pot}}
in addition to an inflaton sector which we leave unspecified. We assume that the possible non-minimal couplings between the field $\chi$ and gravity are so small that they do not affect the following analysis.

Assuming that the mass of the field $\chi$ is much smaller than the Hubble rate during inflation, i.e. $V''\ll H^2$ where $'$ denotes derivative with respect to the field, it will receive excitations from the rapidly expanding background. Perhaps the simplest way of showing this is by making use of the stochastic formalism, which shows that the one-point equilibrium distribution
of the field is
\cite{Starobinsky:1994bd}
\ee{P(\chi) =N \exp\bigg[-\f{8\pi^2}{3H^4}V(\chi)\bigg]\,,\label{eq:p}}
where $N$ is a normalization factor. Specifically, at the end of inflation there will be a non-zero condensate of the $\chi$ field, whose variance at the end of inflation reads 
\ee{
	\label{hstar}
\langle{\chi}_{\rm end}^2\rangle = \sqrt{\frac{3}{2\pi^2}}{\frac{\Gamma(\frac{3}{4})}
		{\Gamma(\frac{1}{4})}} \frac{H_{\rm end}^2}{\sqrt{\lambda}}\approx 0.132 \frac{H_{\rm end}^2}{\sqrt{\lambda}}\,,}
where 'end' refers to the end of inflation and we assumed that the $\chi$ mass term is negligible compared with the interaction term during inflation.
This requires that
\begin{equation}
\label{equ:lambdadom}
    m^2\ll \lambda \langle{\chi}_{\rm end}^2\rangle
    \approx 0.132 \sqrt{\lambda}H_{\rm end}^2.
\end{equation}
Furthermore, the condition that $\chi$ is light during inflation
requires
\begin{eqnarray}
    \langle V''(\chi_{\rm end})\rangle&=&
    m^2+3\lambda \langle{\chi}_{\rm end}^2\rangle
    \nonumber\\\label{eq:lightness}
    &\approx& 
    0.40\sqrt{\lambda}H_{\rm end}^2 \lesssim H_{\rm end}^2.
\end{eqnarray}

The energy density of $\chi$ at the end of inflation is
\begin{equation}
    \rho^{\rm end}_\chi(x)=\frac{\lambda}{4}\chi_{\rm end}(x)^4,
\end{equation}
where we have written the argument $x$ explicitly to highlight the fact that this is a position-dependent quantity.

Initially the $\chi$ field remains frozen and therefore its energy density is constant.
When the Hubble friction drops below the effective mass $V''(\chi)$, the condensate begins to oscillate around the minimum at the origin. Ignoring the bare mass term, this happens when
\ee{
H_{\rm osc}^2=3\lambda \chi_{\rm end}^2
\,;
\qquad  
\frac{a_{\rm osc}}{a_{\rm end}} =\sqrt{\frac{H_{\rm end}}{H_{\rm osc}}}
\,,
\label{eq:osc}}
where 'osc' denotes the instant when the field starts oscillating
and $a_{\rm end}$ is the scale factor at the end of inflation. In the above we have assumed that after inflation the Universe immediately becomes radiation dominated, however extending our analysis to include a reheating phase with an arbitrary equation of state is straightforward, see \cite{Enqvist:2017kzh}.

Assuming that Eq.~(\ref{equ:lambdadom}) is satisfied, the potential is initially dominated by the quartic term, and therefore
the energy density  scales 
 on average as that of radiation, 
 $\propto a^{-4}$ 
 \cite{Kolb:1990vq}
\begin{eqnarray}
\rho_\chi(a)&=&\bigg(\frac{a_{\rm osc}}{a}\bigg)^4\rho_\chi^{\rm end}=\bigg(\frac{a_{\rm osc}}{a}\bigg)^4\frac{\lambda}{4}\chi_{\rm end}^4
\nonumber\\
&=&
\frac{H_{\rm end}^2\chi_{\rm end}^2}{12}\left(\frac{a_{\rm end}}{a}\right)^4
\label{eq:spec}\,,
\end{eqnarray}
where we have used Eq.~(\ref{eq:osc}) to obtain the last expression.
The final dark matter abundance depends on the later evolution of the $\chi$ field, and in the following we consider three scenarios:
(1) the field $\chi$ oscillates coherently until the present day;
(2) it fragments and thermalises with itself, and eventually freezes out while still relativistic; and
(3) it becomes non-relativistic before freezing out.
Which of these scenarios is realised, is determined by the values of the parameters.

\subsection{Coherent oscillations}
\label{sec:coh}
In the simplest case the field $\chi$ simply continues to oscillate in its potential until present day. As the Universe expands, the amplitude of the oscillations decreases, and at
some point the quartic term in (\ref{eq:pot}) will become negligible and the mass term will dominate the evolution 
of the $\chi$ field. After this the energy density of $\chi$ will scale as a cold dark matter component, $\propto a^{-3}$.

We will use the standard approximation where the energy density is assumed to instantaneously go from scaling as $\propto a^{-4}$ to $\propto a^{-3}$.
To calculate when this happens, we obtain the amplitude $\tilde{\chi}$ of the oscillations as
\ee{
\tilde\chi(a)=\frac{a_{\rm osc}}{a}|\chi_{\rm end}|.
}
The onset of dust-like scaling behaviour, denoted with subscript 'dust', is then determined by the condition that the two terms in the potential are equal,
\ee{
\frac{1}{2}m^2 \tilde{\chi}(a_{\rm dust})^2=\frac{\lambda}{4}\tilde{\chi}(a_{\rm dust})^4
~~\Rightarrow~~ 
a_{\rm dust}=
\sqrt{\frac{\lambda}{2}}\frac{|\chi_{\rm end}|}{m}a_{\rm osc}
.\label{eq:m}}
This allows us to write the result for the energy density for $a>a_{\rm dust}$, when the $\chi$ component behaves as dark matter
\ee{\rho_\chi(a)
=
\bigg(\frac{a_{\rm dust}}{a}\bigg)^3\rho_\chi(a_{\rm dust})
=
\sqrt{\frac{\lambda}{8}}\left(\frac{a_{\rm osc}}{a}\right)^3m|\chi_{\rm end}|^3
.\label{eq:final}}
Using this and conservation of entropy in the visible SM sector we can write the energy density at the present time as
\begin{equation}
\label{equ:chi0nofrag}
  \rho_\chi(a_0)  
  =\sqrt{\frac{\lambda}{8}}
  \frac{g_{*S}(T_0)}{g_*(T_{\rm osc})}
    \left(\frac{T_{0}}{T_{\rm osc}}\right)^3
  m|\chi_{\rm end}|^3,
\end{equation}
where $T_0$ and $T_{\rm osc}$ refer to the radiation temperature at the present time and at the start of the $\chi$ oscillations, respectively, $g_{*S}(T_0)\approx 3.909$ is the number of effective entropy degrees of freedom today, and $g_*(T_{\rm osc})=106.75$ is the number of the effective degrees of freedom at the start of the oscillation.

The temperature $T_{\rm osc}$ can be determined from the condition
\begin{equation}
    H_{\rm osc}^2=g_*(T_{\rm osc})\frac{\pi^2 T_{\rm osc}^4}{90\MP^2},
\end{equation}
where $\MP=(8\pi G)^{-1/2}\approx 2.435\times 10^{18}~{\rm GeV}.$
Together with Eq.~(\ref{eq:osc}), it gives
\begin{equation}
\label{equ:Tosc}
    T_{\rm osc}=\left(\frac{270\lambda}{g_*(T_{\rm osc})\pi^2}\chi_{\rm end}^2M_{\rm P}^2\right)^{1/4}.
\end{equation}
Substituting this into Eq.~(\ref{equ:chi0nofrag}) gives
\begin{equation}
    \rho_\chi(a_0)
    =
    \left(\frac{\pi^2}{1080}\right)^{3/4}\frac{g_{*S}(T_0)}{g_*(T_{\rm osc})^{1/4}}
    \frac{mT_0^3}{\lambda^{1/4}}\left(\frac{|\chi_{\rm end}|}{M_{\rm P}}\right)^{3/2}.\label{eq:rocho}
\end{equation}
Note that this is a position-dependent quantity, because $\chi_{\rm end}$ depends on position. Its spatial average can be computed
using the one-point probability distribution (\ref{eq:p}), which gives
\begin{equation}
    \langle|\chi_{\rm end}|^{3/2}\rangle=\frac{6^{3/8}H_{\rm end}^{3/2}}{\lambda^{3/8}\pi^{1/4}\Gamma(1/8)}
    \approx
    \frac{0.1952}{\lambda^{3/8}}H_{\rm end}^{3/2}
    ,
    \label{equ:avgchi32}
\end{equation}
and hence
\begin{equation}
    \langle \rho_\chi(a_0)\rangle
    =\frac{\pi^{5/4}}{6^{15/8}5^{3/4}\Gamma(1/8)}
    \frac{g_{*S}(T_0)}{g_*(T_{\rm osc})^{1/4}}
    \frac{mT_0^3}{\lambda^{5/8}}\left(\frac{H_{\rm end}}{M_{\rm P}}\right)^{3/2}.
\end{equation}
Expressing this an energy fraction $\Omega_\chi$, we can write
\begin{equation}
\label{Omega_chi_oscillations}
\frac{\Omega_\chi h^2}{0.12}
\approx 
\frac{9.37\times 10^6}{\lambda^{5/8}}\left(\frac{H_{\rm end}}{M_{\rm P}}\right)^{3/2}\frac{m}{\rm GeV},
\end{equation}
which should be equal to one for $\chi$ particles to fully account for the observed dark matter abundance.

\subsection{Thermalisation}
\label{sec:ther}
If the coupling $\lambda$ is sufficiently large, the $\chi$ condensate will quickly fragment into $\chi$ particles with finite momenta \cite{Ichikawa:2008ne,Kainulainen:2016vzv}. As discussed in Ref. \cite{Kainulainen:2016vzv} for quartic self-interactions the condition for complete decay of the condensate may be written as
\begin{eqnarray}
\frac{\Gamma(\tilde{\chi}(a_{\rm dec}))}{H_{\rm dec}}&\simeq&\frac{0.023\lambda^{2/3}\tilde{\chi}(a_{\rm dec})}{H_{\rm dec}}=1  \nonumber\\
&\Rightarrow& m<0.023\lambda^{3/2}|\chi_{\rm end}|
\label{eq:therm}\,,
\end{eqnarray}
where $\Gamma(\tilde{\chi})$ is the effective decay rate of the condensate into two $\chi$ particles and we used $\tilde{\chi}(a_{\rm dec})=|\chi_{\rm end}|\sqrt{H_{\rm dec}/H_{\rm osc}}$ and $3\lambda\tilde{\chi}^2(a_{\rm dec})>m^2$ to derive a limit for $m$. If the bare mass was larger than the upper limit, the $\chi$ condensate does not fragment and the result for coherently oscillating condensate \eqref{Omega_chi_oscillations} remains valid. If the condensate does fragment, however, we need to calculate the abundance again.

After thermalisation, the particles will have the temperature
\begin{equation}
\label{chi_temperature}
    T_\chi(a)=\left(\frac{15\lambda}{2\pi^2}\right)^{1/4}\left(\frac{a_{\rm osc}}{a}\right)|\chi_{\rm end}|.
\end{equation}
We will work in the approximation where we assume a sharp transition between the regime with no decay of the condensate and complete thermalisation, as defined by (\ref{eq:therm}). As long as the particles are ultrarelativistic, their energy density will continue to redshift according to Eq. \eqref{eq:spec}, and therefore the exact time of thermalisation does not matter for the following calculation.

To compute the present DM abundance in this case, let us first consider the case where DM freeze-out from the $\chi$ sector heat bath occurs while the particles are still relativistic.
In this case the number density of the $\chi$ particles is simply given by the ultrarelativistic expression,
\begin{eqnarray}
    n_\chi(a)&=&\frac{\zeta(3)}{\pi^2}T_\chi(a)^3 \\ \nonumber
    &=&\left(\frac{15}{2}\right)^{3/4}\frac{\zeta(3)}{\pi^{7/2}}
    \left(\frac{a_{\rm osc}}{a}\right)^3\lambda^{3/4}|\chi_{\rm end}|^3.
\end{eqnarray}
Because after their freeze-out the $\chi$ particles are no longer interacting, this expression remains valid even after they have become non-relativistic. 
The energy density of the $\chi$ particles at the present time is therefore
\begin{align}
\nonumber    \rho_\chi(a_0)&=mn_\chi(a_0)
 \\
    &=\left(\frac{15}{2}\right)^{3/4}\frac{\zeta(3)}{\pi^{7/2}}\nonumber
    \frac{g_{*S}(T_0)}{g_*(T_{\rm osc})}
    \left(\frac{T_{0}}{T_{\rm osc}}\right)^3\lambda^{3/4}m|\chi_{\rm end}|^3
    \nonumber
    \\\label{equ:rhochimassive}
    &= 
    \frac{\zeta(3)}{6\sqrt{6}\pi^2}\frac{g_{*S}(T_0)}{g_*(T_{\rm osc})^{1/4}}
    mT_0^3\left(\frac{|\chi_{\rm end}|}{M_{\rm P}}\right)^{3/2}
\end{align}
where we used Eq.~(\ref{equ:Tosc}).
Using Eq.~(\ref{equ:avgchi32}), the average energy density is therefore
\begin{equation}
    \langle\rho_\chi(a_0)\rangle=\frac{\zeta(3)}{(6\pi^2)^{9/8}\Gamma(1/8)}
     \frac{g_{*S}(T_0)}{g_*(T_{\rm osc})^{1/4}}
    \frac{mT_0^3}{\lambda^{3/8}}
     \left(\frac{H_{\rm end}}{M_{\rm P}}\right)^{3/2}\,.
\end{equation}
Expressed as an energy fraction $\Omega_\chi$, this is
\begin{equation}
\label{Omega_chi_thermalisation}
\frac{\Omega_\chi h^2}{0.12}
\approx 
\frac{2.63\times 10^6}{\lambda^{3/8}}\left(\frac{H_{\rm end}}{M_{\rm P}}\right)^{3/2}\frac{m}{\rm GeV}.
\end{equation}
The difference to Eq. \eqref{Omega_chi_oscillations} is due to thermalisation changing the dependence on $\lambda$, cf. Eq. \eqref{chi_temperature}.

\subsection{Cannibalism}
\label{sec:cann}
In the third scenario, the freeze-out occurs while the DM particles are non-relativistic. In that case, when the self-interactions are large, the $\chi$ particles undergo a phase of \textit{cannibalism}, where the $4\to 2$ self-annihilations dilute the number density and heat up the $\chi$ particles, making their temperature scale in a non-trivial way until the eventual freeze-out \cite{Carlson:1992fn}.

Because entropy is conserved, the ratio $\xi\equiv s_{\rm rad}/s_\chi$, where
$s_{\rm rad}$ and $s_\chi$ denote entropy density of the SM sector and the $\chi$ particle heat bath, respectively, remains constant after the particles have thermalised with each other.
Assuming that the $\chi$ particles thermalise with each other after the fragmentation of the $\chi$ condensate but before the produced particles become non-relativistic, we have
\begin{equation}
\label{xi_rel}
\xi 
= \frac{g_{*S}(T)T^3}{T_\chi^3} = 
36^{3/4}g_*(T_{\rm osc})^{1/4}\left(\frac{M_{\rm P}}{|\chi_{\rm end}|}\right)^{3/2} ,
\end{equation}
where we used Eqs.~(\ref{equ:Tosc}) and (\ref{chi_temperature}). We also assume that the effective entropy and energy degrees of freedom are equal and time-independent at early times.

Then, between the moment when the $\chi$ particles become non-relativistic and their final freeze-out from their internal chemical equilibrium, the ratio of entropy densities is
\begin{equation}
\label{xi_nrel}
\xi =  \frac{s_{\rm rad}}{s^{\rm non-rel}_\chi} =
\frac{2\pi^2g_{*S}(T) T^3}{45xn(x)},
\end{equation}
where $x\equiv m/T_\chi$, and
\begin{equation}
n(x)= \frac{m^3}{(2\pi)^{3/2}}x^{-3/2}e^{-x}
\end{equation}
and 
$s^{\rm non-rel}_\chi = xn(x)$
are the number density and entropy density of a non-relativistic ideal gas, respectively.

By equating Eqs. \eqref{xi_rel} and \eqref{xi_nrel}, we can relate the SM photon temperature to the temperature of $\chi$ particles as
\begin{equation}
\label{TtoTchi}
T = \left(\frac{270\sqrt{6}}{2\pi^2}\right)^{1/3}\frac{g_*(T_{\rm osc})^{1/12}}{g_{*S}(T)^{1/3}}\sqrt{\frac{M_{\rm P}}{\chi_{\rm end}}}n^{1/3}(x)x^{1/3}.
\end{equation}
Therefore, before the DM freeze-out the Hubble parameter can be expressed as
\begin{eqnarray}
\label{hubble}
H &=& \sqrt{\frac{\pi^2g_*(T)}{90}}\frac{T^2}{M_{\rm P}} \\ \nonumber
&=& 
3
\left(\frac{45}{2\pi^2}\right)^{1/6} \left(\frac{g_*(T_{\rm osc})}{g_*(T)}\right)^{1/6}\frac{n(x)^{2/3}x^{2/3}}{\chi_{\rm end}}.
\end{eqnarray}
The $\chi$ particles remain in equilibrium until the Hubble rate and the number density satisfy the freeze-out condition
\begin{equation}
\label{FO_cond}
\frac{\langle \sigma_{4\to 2}v^3\rangle n^3}{H} = 1 ,
\end{equation}
where
in the non-relativistic limit \cite{Tenkanen:2016jic}
\begin{equation}
\langle \sigma_{4\to 2}v^3\rangle \simeq \frac{81\sqrt{3}\lambda^4}{32\pi m^8}.
\end{equation}
The moment of freeze-out is therefore
\begin{equation}
\label{xf}
x_f(\chi_{\rm end}) = \frac{25}{14}W\left(0.1\left(\frac{\lambda^4|\chi_{\rm end}|}{m}\right)^{6/25}\right) ,
\end{equation}
where $W$ is the principal branch of the Lambert W-function and we used Eq. \eqref{hubble} to express $H_{\rm end}$ as a function of $x$. If the ratio \eqref{FO_cond} ever was greater than unity, invoking principle of detailed balance shows that the $\chi$ particles indeed had thermalised with each other prior to their eventual freeze-out. Reminiscent to the standard WIMP case, the final abundance is not sensitive to when thermalisation occurs.

\begin{figure}[t]
	\begin{center}	\includegraphics[width=0.48\textwidth]{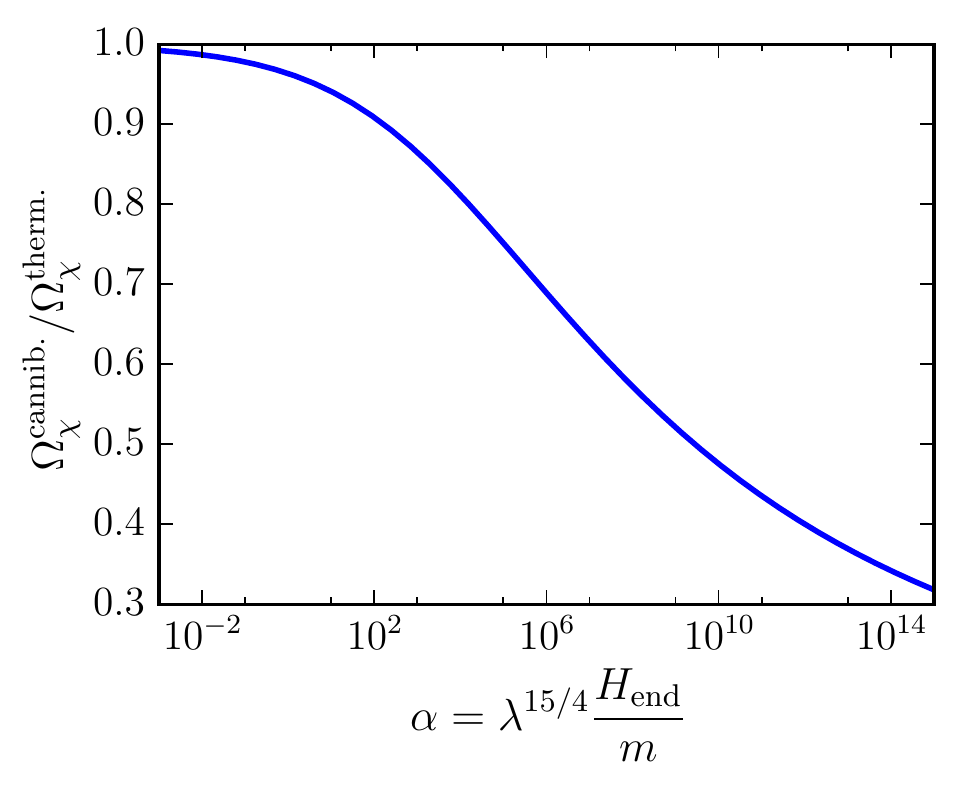}
	\end{center}
 \caption{Ratio of Eq.~(\ref{eq:cannib}) to Eq.~(\ref{Omega_chi_thermalisation}) as a function of $\alpha=\lambda^{15/4}H_{\rm end}/m$, showing the relative suppression of the dark matter abundance due to cannibalism.}
    \label{fig:omegaratio}
\end{figure}

It is then straightforward to compute the present abundance of $\chi$ particles. Expressing the radiation temperature at the time of freeze-out by $T_f$, we have
\begin{eqnarray}
\nonumber \rho_\chi(a_0) &=& mn(x_f)\frac{g_{*S}(T_0)T_0^3}{g_{*S}(T_f)T_f^3}
\\
&=&
\frac{\pi^2}{135\sqrt{6}}\frac{g_{*S}(T_0)}{g_*(T_{\rm osc})^{1/4}}\frac{mT_0^3}{x_f(\chi_{\rm end})}
\left(\frac{|\chi_{\rm end}|}{M_{\rm P}}\right)^{3/2},
\label{eq:r1}
\end{eqnarray}
where $T_0$ is the CMB photon temperature today. This expression is valid when $x_f\gtrsim 1$,
whereas in the limit $x_f\ll 1$, the density is given by Eq.~(\ref{equ:rhochimassive}).
To cover the whole range of $\chi_{\rm end}$, we therefore interpolate between them with
\begin{align}
    \rho_\chi(a_0) \simeq
\frac{\pi^2}{135\sqrt{6}}\frac{g_{*S}(T_0)}{g_*(T_{\rm osc})^{1/4}}\frac{mT_0^3}{X_f(\chi_{\rm end})}
\left(\frac{|\chi_{\rm end}|}{M_{\rm P}}\right)^{3/2},
\end{align}
where
\begin{align}
    X_f(\chi_{\rm end})=x_f(\chi_{\rm end})+\frac{2\pi^4}{45\zeta(3)}\approx x_f(\chi_{\rm end})+3.602.
\end{align}
Thus
\begin{equation}
\frac{\Omega_\chi h^2}{0.12} \simeq \frac{1.56\times 10^{8}}{g_{*}^{1/4}(T_{\rm osc})M_{\rm P}^{3/2}}
\left\langle\frac{|\chi_{\rm end}|^{3/2}}{X_f(\chi_{\rm end})}\right\rangle
\frac{m}{\rm GeV} ,\label{eq:cannib}
\end{equation}
where the expectation value needs to be computed numerically.
In Fig.~\ref{fig:omegaratio} we show the ratio of Eq.~(\ref{eq:cannib}) to Eq.~(\ref{Omega_chi_thermalisation}), which shows the relative suppression of the dark matter abundance due to cannibalism.
It is easy to see by scaling that the ratio only depends on the parameter combination $\alpha=\lambda^{15/4}H_{\rm end}/m$, and approaches one when $\alpha\rightarrow 0$.

\section{Isocurvature perturbations}
\label{sec:isocurvature}

As the dark matter energy density is position-dependent and does not necessarily track the one of baryonic matter at all scales, we need to worry about observational constraints on isocurvature perturbations. In order for the $\chi$ field generated during inflation to be a viable dark matter candidate it must not violate the current rather stringent observational bounds from the Planck satellite allowing only a small isocurvature component \cite{Akrami:2018odb}.

Isocurvature between two components is defined as
\ee{
\label{isocurvature2}
S\equiv 3H\frac{\delta\rho_{i}}{\dot{\rho_{i}}}-3H\frac{\delta\rho_{j}}{\dot{\rho_{j}}}\,,
}
which straightforwardly gives the isocurvature between CDM and radiation 
\ee{
\label{isocurvature3}
S\equiv \frac{\delta\rho_\chi}{{\rho_\chi}}-\frac{3}{4}\frac{\delta\rho_{\gamma}}{{\rho_{\gamma}}}\,
}
at late times.

For calculating the isocurvature perturbations we choose a gauge where there are no fluctuations in the inflaton field. In this gauge the curvature perturbation then manifests as a perturbation in the scale factor ${\delta\rho_{\gamma}}/{\rho_{\gamma}}=-4{\delta a}/{a}$ and furthermore the isocurvature in the DM fluid $\rho_\chi$ is only sourced by the perturbations of $\chi_{\rm end}$. 
In each of the three scenarios considered in Section~\ref{sec:CDM}, we can write
$\rho_\chi(a_0)\propto f(\chi_{\rm end})$. For Eqs.~(\ref{Omega_chi_oscillations}) and
(\ref{Omega_chi_thermalisation}), we can choose $f(\chi_{\rm end})=|\chi_{\rm end}|^{3/2}$, and for Eq.~(\ref{eq:cannib}),
\begin{equation}
    f(\chi_{\rm end})=\frac{|\chi_{\rm end}|^{3/2}}{X_f(\chi_{\rm end})}.
\end{equation}
The isocurvature perturbations (\ref{isocurvature3}) are then given by
\ee{
	S=\frac{\delta(f(\chi_{\rm end}))}{\langle f(\chi_{\rm end})\rangle}\,,\label{isocurvature4}
}
where the perturbation is defined as
\ee{\delta\left(f(\chi)\right)\equiv
f(\chi(x)) -\langle f(\chi)\rangle\,,
\label{eq:fluct}}
and from now on for simplicity we drop the subscript 'end'. So quite naturally, if the $\chi$ field is perfectly homogeneous after inflation, isocurvature strictly vanishes at late times. In our case, as shown by Eqs. (\ref{eq:p}) and (\ref{hstar}), the field is light during inflation and hence there are fluctuations in $\chi$, which is then a genuine isocurvature component. Importantly, despite the field $\chi$ having fluctuations, it has a vanishing one-point function.

Eqs.  (\ref{isocurvature4}) and (\ref{eq:fluct}) then give the two-point correlator for the isocurvature as a function of the $n$-point correlators of the field $\chi$,
\ee{
\langle S(0)S(r) \rangle =\frac{\langle f(\chi(0))f(\chi(r))\rangle-\langle f(\chi)\rangle^2}{\langle f(\chi)\rangle^2}\,.\label{eq:disc}
}
Equation (\ref{eq:disc}) is an equal-time correlator between two different points in space. Using de Sitter invariance, it can be computed as an analytic continuation of an unequal-time correlator. Generalising the analysis of Ref.~\cite{Starobinsky:1994bd}, one can write an unequal-time correlator in terms of the spectral expansion as\footnote{Note that our normalisation of the eigenvalues $\Lambda_n$ differs from Ref. \cite{Starobinsky:1994bd} by factor $H$.}
\ee{
\frac{\langle f(\chi(0))f(\chi(t))\rangle}{\langle f(\chi)\rangle^2}
=\sum_nf_n^2 e^{-\Lambda_n Ht}
\,,
\label{equ:spectral}
}
where
\begin{equation}
f_n=\frac{\int d \chi \psi_0(\chi)f(\chi)\psi_n(\chi)}{\int d \chi \psi_0(\chi)f(\chi)\psi_0(\chi)}
\,,
\end{equation}
and $\Lambda_n$ and $\psi_n$ are the eigenvalues and orthonormal eigenvectors, respectively, of the eigenvalue equation
\begin{equation}
\bigg[\frac{1}{2}\frac{\partial^2}{\partial \chi^2}-\frac{1}{2}\left(v'( \chi)^2-v''( \chi)\right)\bigg]\psi_n( \chi)
=-\frac{4\pi^2\Lambda_n}{H^2}\psi_n( \chi)\,,\label{e:sch}
\end{equation} 
with 
\begin{equation}
    v( \chi)=\frac{4\pi^2}{3H^4}V( \chi).
\end{equation}

In our case, $f(\chi)$ is an even function, and therefore only even eigenvalues contribute to the spectral expansion. Furthermore, $\Lambda_0=0$, so the
$n=0$ term cancels the disconnected part of the correlator. The leading non-trivial term at large $t$ is therefore to a good accuracy
\begin{equation}
\frac{\langle f(0)f(t)\rangle-\langle f\rangle^2 }{\langle f\rangle^2}
\approx
f_2^2 e^{-\Lambda_2Ht}.
\label{equ:spectralapprox}
\end{equation}

\begin{figure}[t]
	\begin{center}	\includegraphics[width=0.48\textwidth]{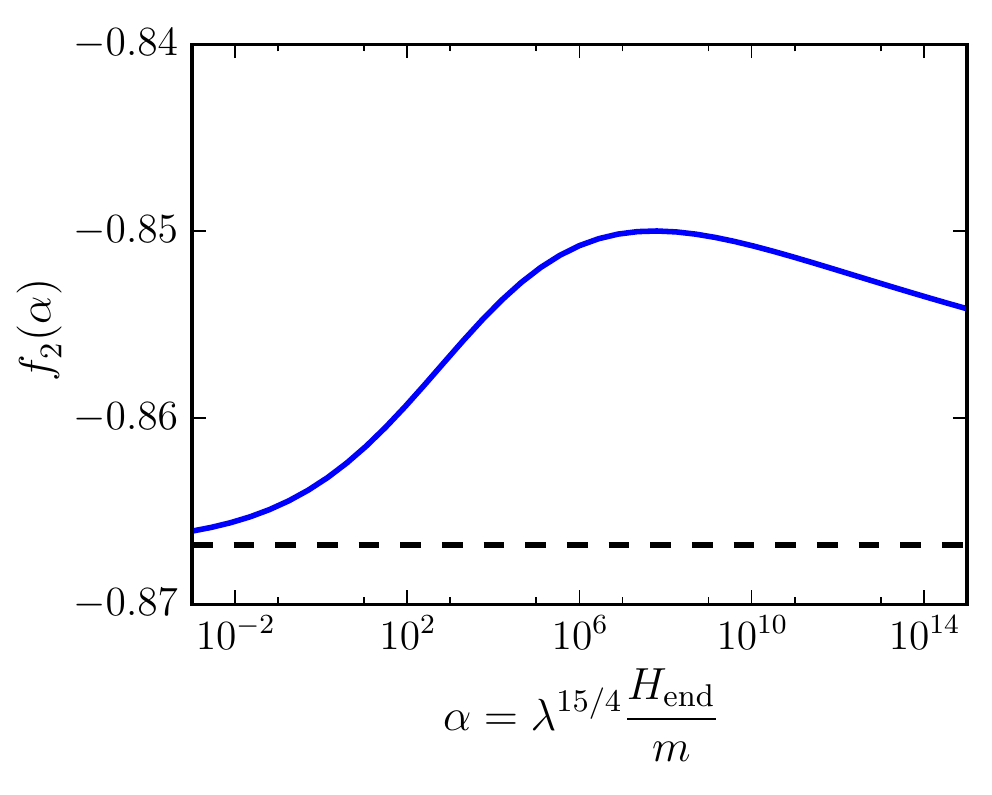}
	\end{center}
	\caption{The dependence of $f_2$ on $\alpha=\lambda^{15/4}H_{\rm end}/m$ \label{fig:f2}, leading to depletion of the DM abundance via cannibalism, as discussed in Sec. \ref{sec:cann}. The dashed line corresponds to the case (\ref{eq:f2}) with no cannibalism.}
\end{figure}

A numerical solution of the eigenvalue equation (\ref{e:sch}) gives~\cite{Starobinsky:1994bd}
\begin{equation}
    \Lambda_2\approx 4.45370\sqrt{\frac{\lambda}{24\pi^2}}
    \approx 0.28938\lambda^{1/2}.
\end{equation}
Again, the value of $f_2$ depends on the parameters only through the combination $\alpha=\lambda^{15/4}H_{\rm end}/m$,
and $\alpha=0$ corresponds to the case with no cannibalism. In this limit, we obtain
\begin{equation}
f_2\approx -0.86683\,,\label{eq:f2}
\end{equation}
When cannibalism does occur, the value of  
$f_2$ varies by only a few percent as shown in Fig. \ref{fig:f2}. For this reason a very good approximation is to take $f_2$ a constant given by (\ref{eq:f2}), which we will choose for now on.

Making use of de Sitter invariance we may relate the unequal-time correlator (\ref{equ:spectral}) to an equal-time correlator with spatial separation by writing $t\rightarrow (2/H)\ln(aHr)$ and the isocurvature correlator (\ref{eq:disc}) can then be expressed as
\begin{equation}
\langle S(0)S(r) \rangle 
\approx
f_2^2(aHr)^{-2\Lambda_2}\label{eq:num}\,,
\end{equation}
where $a$ and $H$ are to be evaluated at the end of inflation.

The power spectrum ${\cal P}(k)$ of the isocurvature perturbations is defined in the standard manner as a Fourier transform
\ee{
{\cal P}(k)=\frac{k^3}{2\pi^2}\int d^3x e^{i\vec{k}\cdot\vec{x}}\langle S(0)S(\vec{x})\rangle
\label{eq:P}\,.
}
Upon substituting Eq.~(\ref{eq:num}), we obtain
\ee{
\mathcal{P}(k)\approx\mathcal{A}\bigg(\frac{k}{\tilde{k}}\bigg)^{2\Lambda_2}\,,\label{eq:pp}
} where the scale $\tilde{k}=a_{\rm end}H_{\rm end}$ is the horizon scale at the end of inflation
and
\begin{eqnarray}
\mathcal{A} &=&
\frac{2f_2^2}{\pi}\Gamma\left(2-2\Lambda_2\right)\sin \left(\Lambda_2\pi\right)
\nonumber\\
&=& 0.4349\sqrt{\lambda}+O\left(\lambda\right).
\end{eqnarray}
Because $\Lambda_2>0$, the spectrum is blue. The spectral index, defined as
$\mathcal{P}(k)\propto k^{n-1}$, is
\begin{equation}
    n-1=2\Lambda_2\approx 0.5788 \sqrt{\lambda}.
    \label{equ:spectralinex}
\end{equation}

The difference between the pivot scale $k_*=0.05 {\rm Mpc}^{-1}$ at which the isocurvature perturbations are measured and the horizon scale at the end of inflation $\tilde{k}$ can be characterized with the $e$-fold number
\begin{equation}
N_*\equiv
    \ln\left(\frac{\tilde{k}}{k_*}\right) \approx 56 + \frac12\ln\left(\frac{H_{\rm end}}{8\times 10^{13} {\rm GeV}}\right) ,
\end{equation}
where, in the last expression, we have neglected the small change in $H$ during inflation and maintained our assumption that the Universe is radiation-dominated from the end of inflation. Hence our final expression for the isocurvature spectrum at the scale $k_*$
is
\begin{equation}
{\cal P}(k_*)
= {\cal A}e^{-2\Lambda_2N_*}
\approx 0.43\sqrt{\lambda}e^{-0.58\sqrt{\lambda}N_*}.
\label{equ:powerspectrum}
\end{equation}

\section{Results}
\label{results}

\begin{figure}[t]
	\begin{center}	\includegraphics[width=0.47\textwidth]{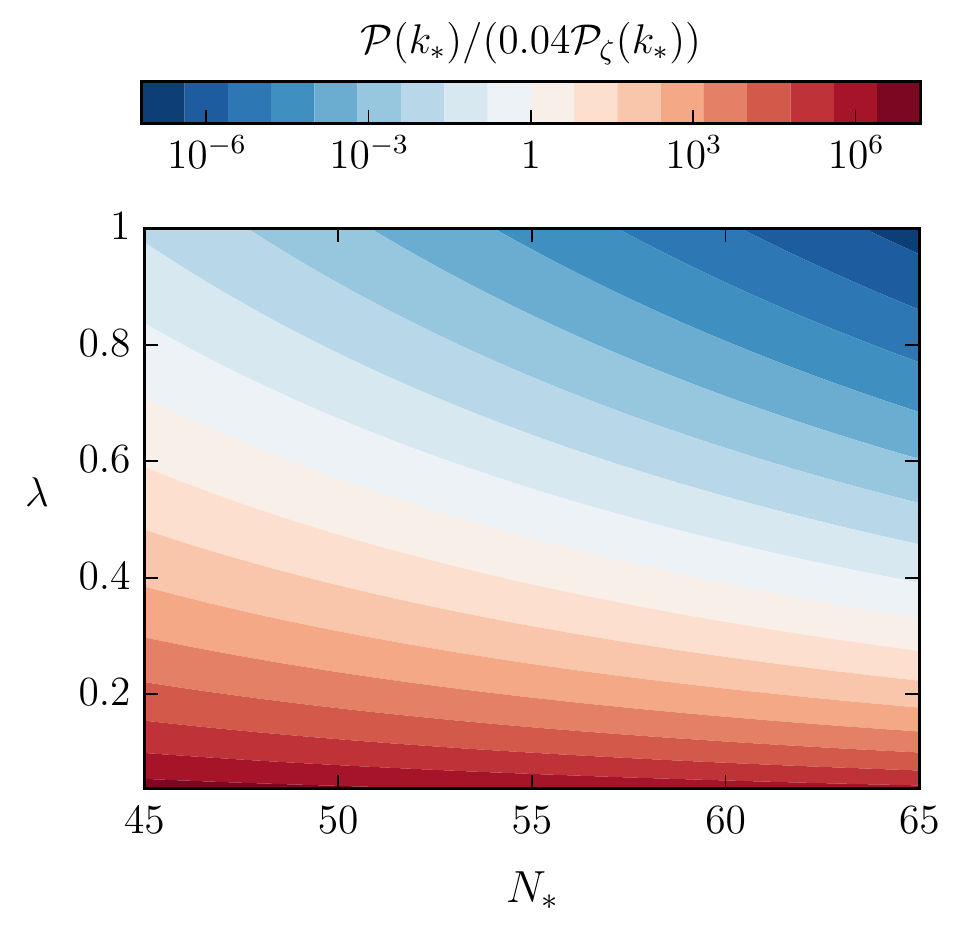}
	\end{center}
	\caption{Isocurvature contours in $(N_*,\lambda)$ space, placing a lower limit on $\lambda$. \label{fig:cont}}
\end{figure}

The current bound for uncorrelated isocurvature between DM and the CMB photons can be expressed as a fraction of the curvature power spectrum ${\cal P}_\zeta =2.2\times 10^{-9}$ as \cite{Akrami:2018odb}
\ee{\label{eq:isob}
\mathcal{P}(k_*)\lesssim 0.040\mathcal{P}_{\zeta}(k_*)\,.
}
Combining this with Eq.~(\ref{equ:powerspectrum}) and choosing $N_*=56$
gives the constraint\footnote{For completeness, we note that there is also another branch of solutions at $\lambda\lesssim 10^{-19}$. However, as this regime is phenomenologically less interesting, in the present paper we neglect this possibility.}
\begin{equation}
    \lambda\gtrsim 0.45.
\label{equ:lambdabound}
\end{equation}
In Figure \ref{fig:cont}, we show the isocurvature contours in $(N_*,\lambda)$ space, which demonstrate that for all $N_*\in (45,65)$ the isocurvature bound (\ref{eq:isob}) can be avoided while maintaining a perturbative self-interaction.

\begin{figure}[t]
	\begin{center}
	\includegraphics[width=0.48\textwidth]{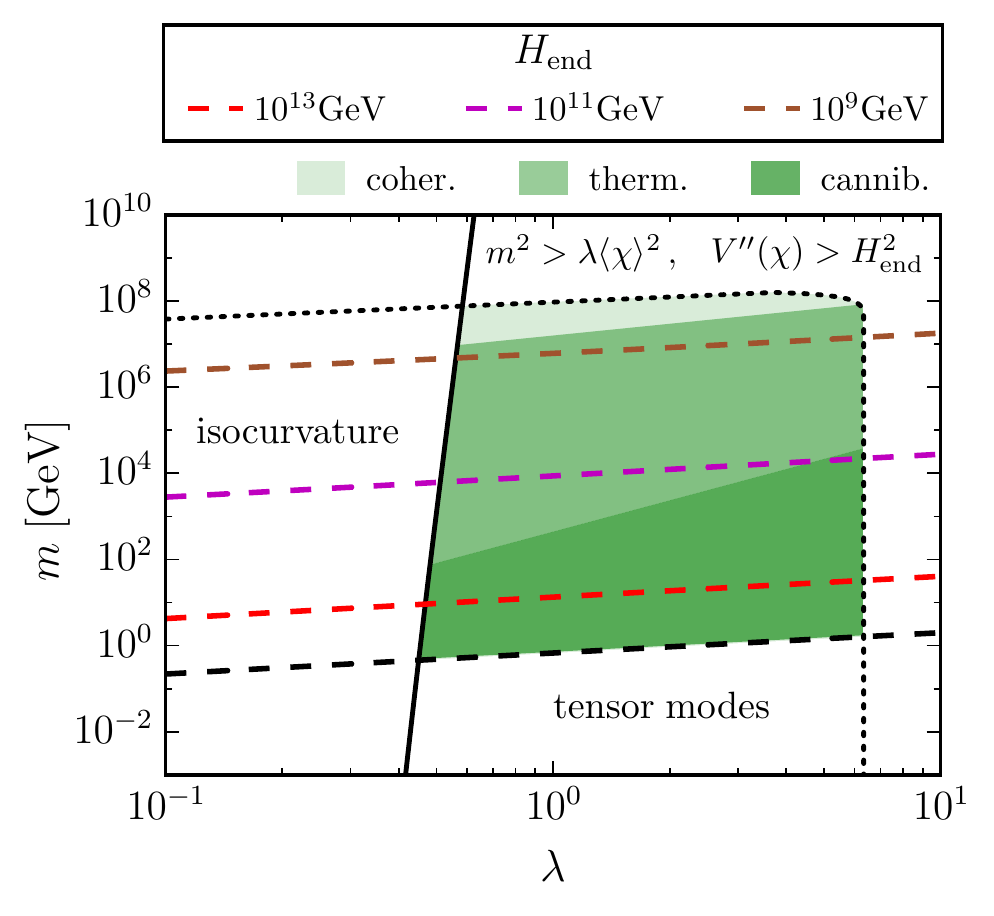}
	\end{center}
	\caption{The green region shows the allowed values of the parameters $m$ and $\lambda$, bounded by 
	Eqs.~{(\ref{equ:lambdadom}), (\ref{eq:lightness})}, {(\ref{equ:powerspectrum})} and (\ref{equ:Hbound}),
	with different shades of green corresponding to the three scenarios discussed in Section~\ref{sec:CDM}. 
	The coloured dashed contours show the dependence on the inflationary scale $H_{\rm end}$.\label{fig:lamvm}}
\end{figure}

In Figure \ref{fig:lamvm}, we show the allowed parameter space for the model. The shown bounds are given by requiring that the $\chi$ potential is quartic and the field is light during inflation, that the isocurvature spectrum is below the Planck constraint, and that the scale of inflation is below the bound given by the non-observation of tensor modes. These are given, respectively, 
by Eqs.~{(\ref{equ:lambdadom}), (\ref{eq:lightness})}, {(\ref{equ:powerspectrum})} and \cite{Ade:2018gkx}
\begin{equation}
\label{equ:Hbound}
H_{\rm end}\lesssim 8\times 10^{13}~{\rm GeV}.
\end{equation}
The different shades of green denote, from the lightest to darkest colour, the regions where the condensate never fragments but oscillates coherently, where the field fragments and the produced particles thermalise but freeze-out while still relativistic, and finally where cannibalism may take place. For details, see Section \ref{sec:CDM}. The borderline for the region with cannibalism is given by setting $x_f=3.60$ in Eq. (\ref{xf}) making Eq.~(\ref{equ:rhochimassive}) equal to Eq.~(\ref{eq:r1}), and the border between the case of a coherently oscillating condensate and the case of thermalisation may be solved from Eq. (\ref{eq:therm}). For simplicity, for the isocurvature bound we have assumed that the $\chi$ condensate fragments throughout the parameter space.
We see that the scenario works for a broad range of masses $1~{\rm GeV}\lesssim m\lesssim 10^8~{\rm GeV}$, depending on the scale of inflation. The lower bound for the quartic DM self-coupling is $\lambda\gtrsim 0.45$, as discussed above.

Finally, we discuss observational properties of the scenario. In the near future, the primordial tensor-to-scalar ratio $r$ can be either detected above or constrained at the level $r\sim 10^{-3}$ \cite{Matsumura:2013aja,Wu:2016hul,Abazajian:2016yjj,Ade:2018sbj}, which corresponds to $H_{\rm end}\lesssim 8\times 10^{12}$ GeV. On the other hand, future observations of the CMB and the large scale structure of the Universe will improve limits on primordial isocurvature or, in the best possible scenario, detect it \cite{Abazajian:2016yjj,Ade:2018sbj}. Furthermore, the Bullet Cluster and collisions between other galaxy clusters can be used to place an upper bound on the self-interaction cross section over DM mass, $\sigma/m\le 1$cm$^2$g$^{-1} \approx 4.6\times 10^3{\rm GeV}^{-3}$ ~\cite{Markevitch:2003at,Randall:2007ph,Rocha:2012jg,Peter:2012jh,Harvey:2015hha}, which is relevant for small $m$. For our theory \cite{Heikinheimo:2016yds}
\begin{equation}
    \frac{\sigma}{m} = \frac{9\lambda^2}{32\pi m^3} ,
\end{equation}
so the the Bullet Cluster imposes a constraint
\beq
\label{eq:sigmaDMbound}
\frac{m}{\rm GeV} > 0.027\left(\frac{\sigma/m}{{\rm cm}^2/{\rm g}}\right)^{-1/3} \lambda^{2/3}\,.
\eeq
This bound, however, is weaker than the other constraints discussed above. At the same time it shows that if sizeable DM self-interactions $\sigma/m\gtrsim 10^{-4}$cm$^2$g$^{-1}$ are discovered in the future, that would rule out the simplest scenario considered in this paper where all of DM is generated by inflationary fluctuations with a typical spectrum.

\begin{figure}
	\begin{center}
	\includegraphics[width=0.48\textwidth]{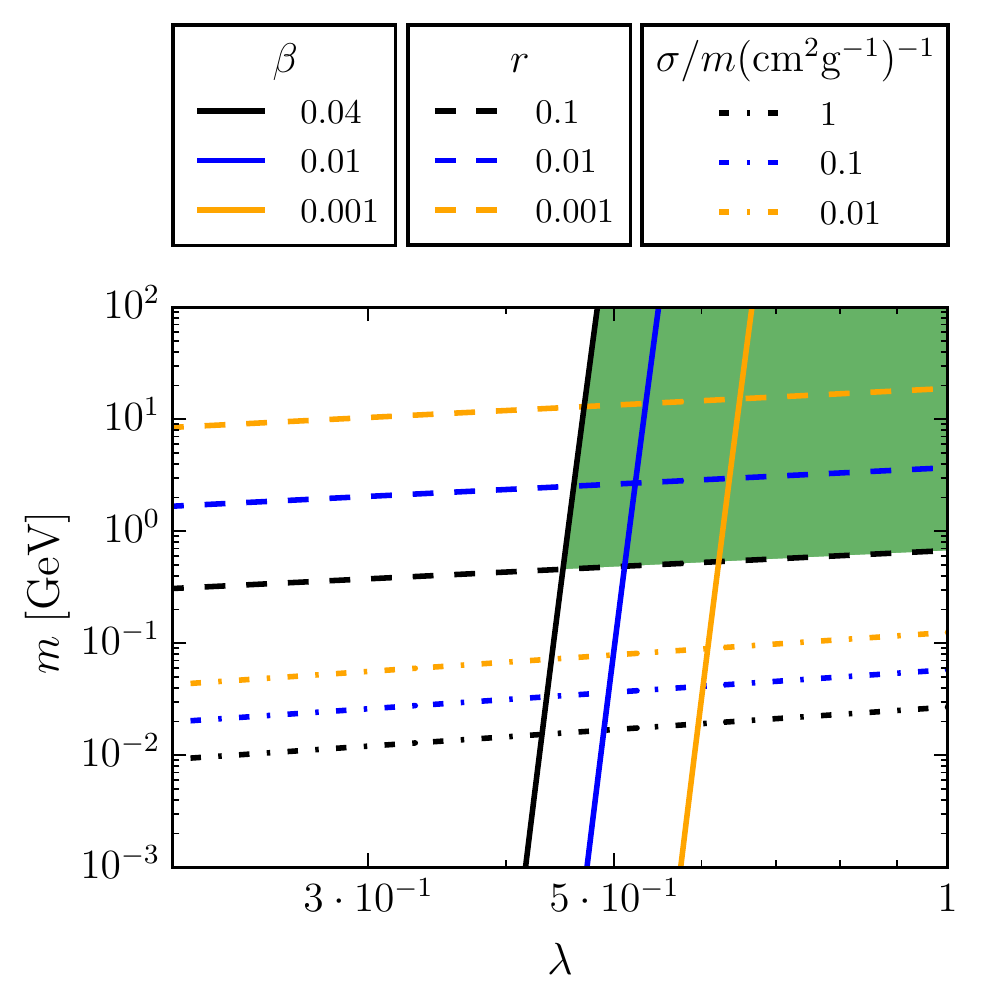}
	\end{center}
	\caption{
	A close up of the parameter space in Fig. \ref{fig:lamvm}. The solid lines denote forecasts for the isocurvature parameter $\beta\simeq \mathcal{P}(k_*)/\mathcal{P}_\zeta(k_*)$ as marked in the legend, dashed lines denote forecasts for the tensor-to-scalar ratio $r$, and dot-dashed lines for the DM self-interaction cross section over DM mass $\sigma/m$.}
	\label{fig:lamvmzoom}
\end{figure}

Detection of either $\beta$ or $r$ would constrain the parameters of the model, and discovery of both primordial isocurvature perturbations and B-mode polarization in the CMB would single out a {\it point} in the model parameter space. As this would unavoidably be at small values of $m$ (as can be seen in Fig. \ref{fig:lamvmzoom}), confirmed detection of non-zero DM self-interactions should indeed provide an additional way to probe the scenario, either validating or ruling out the model studied in this paper.

\section{Conclusions}
\label{conclusions}

We have shown that a decoupled sector consisting of a single massive self-interacting scalar that interacts with the Standard Model only gravitationally is a viable candidate for DM and requires no fine-tuning, providing arguably one of the simplest DM models to date. Our analysis can be straightforwardly generalised to scenarios in which the scalar interacts with other hidden sector fields or (sufficiently weakly) with the Standard Model. As discussed above, the analysis can also be easily modified to accommodate other cosmological histories. The analysis and the resulting bounds are therefore expected to be generic to most weakly coupled DM models with scalar fields which were sufficiently light during inflation.

\section*{Acknowledgements}
We thank David E. Kaplan and Stephen Stopyra for discussions.
T.M. and A.R. are supported by the U.K. Science and Technology Facilities Council grant ST/P000762/1. T.M. is also supported by the Estonian Research Council via the Mobilitas Plus grant MOBJD323. T.T. is supported by the Simons Foundation and the U.K. Science and Technology Facilities Council grant ST/J001546/1.

\bibliography{DM_isocurvature}

\begin{thebibliography}{46}%
\makeatletter
\providecommand \@ifxundefined [1]{%
 \@ifx{#1\undefined}
}%
\providecommand \@ifnum [1]{%
 \ifnum #1\expandafter \@firstoftwo
 \else \expandafter \@secondoftwo
 \fi
}%
\providecommand \@ifx [1]{%
 \ifx #1\expandafter \@firstoftwo
 \else \expandafter \@secondoftwo
 \fi
}%
\providecommand \natexlab [1]{#1}%
\providecommand \enquote  [1]{``#1''}%
\providecommand \bibnamefont  [1]{#1}%
\providecommand \bibfnamefont [1]{#1}%
\providecommand \citenamefont [1]{#1}%
\providecommand \href@noop [0]{\@secondoftwo}%
\providecommand \href [0]{\begingroup \@sanitize@url \@href}%
\providecommand \@href[1]{\@@startlink{#1}\@@href}%
\providecommand \@@href[1]{\endgroup#1\@@endlink}%
\providecommand \@sanitize@url [0]{\catcode `\\12\catcode `\$12\catcode
  `\&12\catcode `\#12\catcode `\^12\catcode `\_12\catcode `\%12\relax}%
\providecommand \@@startlink[1]{}%
\providecommand \@@endlink[0]{}%
\providecommand \url  [0]{\begingroup\@sanitize@url \@url }%
\providecommand \@url [1]{\endgroup\@href {#1}{\urlprefix }}%
\providecommand \urlprefix  [0]{URL }%
\providecommand \Eprint [0]{\href }%
\providecommand \doibase [0]{http://dx.doi.org/}%
\providecommand \selectlanguage [0]{\@gobble}%
\providecommand \bibinfo  [0]{\@secondoftwo}%
\providecommand \bibfield  [0]{\@secondoftwo}%
\providecommand \translation [1]{[#1]}%
\providecommand \BibitemOpen [0]{}%
\providecommand \bibitemStop [0]{}%
\providecommand \bibitemNoStop [0]{.\EOS\space}%
\providecommand \EOS [0]{\spacefactor3000\relax}%
\providecommand \BibitemShut  [1]{\csname bibitem#1\endcsname}%
\let\auto@bib@innerbib\@empty
\bibitem [{\citenamefont {Bertone}\ and\ \citenamefont
  {Hooper}(2018)}]{Bertone:2016nfn}%
  \BibitemOpen
  \bibfield  {author} {\bibinfo {author} {\bibfnamefont {G.}~\bibnamefont
  {Bertone}}\ and\ \bibinfo {author} {\bibfnamefont {D.}~\bibnamefont
  {Hooper}},\ }\href {\doibase 10.1103/RevModPhys.90.045002} {\bibfield
  {journal} {\bibinfo  {journal} {Rev. Mod. Phys.}\ }\textbf {\bibinfo {volume}
  {90}},\ \bibinfo {pages} {045002} (\bibinfo {year} {2018})},\ \Eprint
  {http://arxiv.org/abs/1605.04909} {arXiv:1605.04909 [astro-ph.CO]}
  \BibitemShut {NoStop}%
\bibitem [{\citenamefont {Aghanim}\ \emph {et~al.}(2018)\citenamefont {Aghanim}
  \emph {et~al.}}]{Aghanim:2018eyx}%
  \BibitemOpen
  \bibfield  {author} {\bibinfo {author} {\bibfnamefont {N.}~\bibnamefont
  {Aghanim}} \emph {et~al.} (\bibinfo {collaboration} {Planck}),\ }\href@noop
  {} {\  (\bibinfo {year} {2018})},\ \Eprint {http://arxiv.org/abs/1807.06209}
  {arXiv:1807.06209 [astro-ph.CO]} \BibitemShut {NoStop}%
\bibitem [{\citenamefont {Arcadi}\ \emph {et~al.}(2018)\citenamefont {Arcadi},
  \citenamefont {Dutra}, \citenamefont {Ghosh}, \citenamefont {Lindner},
  \citenamefont {Mambrini}, \citenamefont {Pierre}, \citenamefont {Profumo},\
  and\ \citenamefont {Queiroz}}]{Arcadi:2017kky}%
  \BibitemOpen
  \bibfield  {author} {\bibinfo {author} {\bibfnamefont {G.}~\bibnamefont
  {Arcadi}}, \bibinfo {author} {\bibfnamefont {M.}~\bibnamefont {Dutra}},
  \bibinfo {author} {\bibfnamefont {P.}~\bibnamefont {Ghosh}}, \bibinfo
  {author} {\bibfnamefont {M.}~\bibnamefont {Lindner}}, \bibinfo {author}
  {\bibfnamefont {Y.}~\bibnamefont {Mambrini}}, \bibinfo {author}
  {\bibfnamefont {M.}~\bibnamefont {Pierre}}, \bibinfo {author} {\bibfnamefont
  {S.}~\bibnamefont {Profumo}}, \ and\ \bibinfo {author} {\bibfnamefont
  {F.~S.}\ \bibnamefont {Queiroz}},\ }\href {\doibase
  10.1140/epjc/s10052-018-5662-y} {\bibfield  {journal} {\bibinfo  {journal}
  {Eur. Phys. J.}\ }\textbf {\bibinfo {volume} {C78}},\ \bibinfo {pages} {203}
  (\bibinfo {year} {2018})},\ \Eprint {http://arxiv.org/abs/1703.07364}
  {arXiv:1703.07364 [hep-ph]} \BibitemShut {NoStop}%
\bibitem [{\citenamefont {Kolb}\ and\ \citenamefont
  {Turner}(1990)}]{Kolb:1990vq}%
  \BibitemOpen
  \bibfield  {author} {\bibinfo {author} {\bibfnamefont {E.~W.}\ \bibnamefont
  {Kolb}}\ and\ \bibinfo {author} {\bibfnamefont {M.~S.}\ \bibnamefont
  {Turner}},\ }\href@noop {} {\bibfield  {journal} {\bibinfo  {journal} {Front.
  Phys.}\ }\textbf {\bibinfo {volume} {69}},\ \bibinfo {pages} {1} (\bibinfo
  {year} {1990})}\BibitemShut {NoStop}%
\bibitem [{\citenamefont {McDonald}(2002)}]{McDonald:2001vt}%
  \BibitemOpen
  \bibfield  {author} {\bibinfo {author} {\bibfnamefont {J.}~\bibnamefont
  {McDonald}},\ }\href {\doibase 10.1103/PhysRevLett.88.091304} {\bibfield
  {journal} {\bibinfo  {journal} {Phys.Rev.Lett.}\ }\textbf {\bibinfo {volume}
  {88}},\ \bibinfo {pages} {091304} (\bibinfo {year} {2002})},\ \Eprint
  {http://arxiv.org/abs/hep-ph/0106249} {arXiv:hep-ph/0106249 [hep-ph]}
  \BibitemShut {NoStop}%
\bibitem [{\citenamefont {Hall}\ \emph {et~al.}(2010)\citenamefont {Hall},
  \citenamefont {Jedamzik}, \citenamefont {March-Russell},\ and\ \citenamefont
  {West}}]{Hall:2009bx}%
  \BibitemOpen
  \bibfield  {author} {\bibinfo {author} {\bibfnamefont {L.~J.}\ \bibnamefont
  {Hall}}, \bibinfo {author} {\bibfnamefont {K.}~\bibnamefont {Jedamzik}},
  \bibinfo {author} {\bibfnamefont {J.}~\bibnamefont {March-Russell}}, \ and\
  \bibinfo {author} {\bibfnamefont {S.~M.}\ \bibnamefont {West}},\ }\href
  {\doibase 10.1007/JHEP03(2010)080} {\bibfield  {journal} {\bibinfo  {journal}
  {JHEP}\ }\textbf {\bibinfo {volume} {1003}},\ \bibinfo {pages} {080}
  (\bibinfo {year} {2010})},\ \Eprint {http://arxiv.org/abs/0911.1120}
  {arXiv:0911.1120 [hep-ph]} \BibitemShut {NoStop}%
\bibitem [{\citenamefont {Bernal}\ \emph {et~al.}(2017)\citenamefont {Bernal},
  \citenamefont {Heikinheimo}, \citenamefont {Tenkanen}, \citenamefont
  {Tuominen},\ and\ \citenamefont {Vaskonen}}]{Bernal:2017kxu}%
  \BibitemOpen
  \bibfield  {author} {\bibinfo {author} {\bibfnamefont {N.}~\bibnamefont
  {Bernal}}, \bibinfo {author} {\bibfnamefont {M.}~\bibnamefont {Heikinheimo}},
  \bibinfo {author} {\bibfnamefont {T.}~\bibnamefont {Tenkanen}}, \bibinfo
  {author} {\bibfnamefont {K.}~\bibnamefont {Tuominen}}, \ and\ \bibinfo
  {author} {\bibfnamefont {V.}~\bibnamefont {Vaskonen}},\ }\href {\doibase
  10.1142/S0217751X1730023X} {\bibfield  {journal} {\bibinfo  {journal} {Int.
  J. Mod. Phys.}\ }\textbf {\bibinfo {volume} {A32}},\ \bibinfo {pages}
  {1730023} (\bibinfo {year} {2017})},\ \Eprint
  {http://arxiv.org/abs/1706.07442} {arXiv:1706.07442 [hep-ph]} \BibitemShut
  {NoStop}%
\bibitem [{\citenamefont {Ford}(1987)}]{Ford:1986sy}%
  \BibitemOpen
  \bibfield  {author} {\bibinfo {author} {\bibfnamefont {L.~H.}\ \bibnamefont
  {Ford}},\ }\href {\doibase 10.1103/PhysRevD.35.2955} {\bibfield  {journal}
  {\bibinfo  {journal} {Phys. Rev.}\ }\textbf {\bibinfo {volume} {D35}},\
  \bibinfo {pages} {2955} (\bibinfo {year} {1987})}\BibitemShut {NoStop}%
\bibitem [{\citenamefont {Kolb}\ \emph {et~al.}(1999)\citenamefont {Kolb},
  \citenamefont {Chung},\ and\ \citenamefont {Riotto}}]{Kolb:1998ki}%
  \BibitemOpen
  \bibfield  {author} {\bibinfo {author} {\bibfnamefont {E.~W.}\ \bibnamefont
  {Kolb}}, \bibinfo {author} {\bibfnamefont {D.~J.~H.}\ \bibnamefont {Chung}},
  \ and\ \bibinfo {author} {\bibfnamefont {A.}~\bibnamefont {Riotto}},\ }\href
  {\doibase 10.1063/1.59655} {\bibfield  {journal} {\bibinfo  {journal} {AIP
  Conf. Proc.}\ }\textbf {\bibinfo {volume} {484}},\ \bibinfo {pages} {91}
  (\bibinfo {year} {1999})},\ \bibinfo {note} {[592(1999)]},\ \Eprint
  {http://arxiv.org/abs/hep-ph/9810361} {arXiv:hep-ph/9810361 [hep-ph]}
  \BibitemShut {NoStop}%
\bibitem [{\citenamefont {Chung}\ \emph {et~al.}(1999)\citenamefont {Chung},
  \citenamefont {Kolb},\ and\ \citenamefont {Riotto}}]{Chung:1998zb}%
  \BibitemOpen
  \bibfield  {author} {\bibinfo {author} {\bibfnamefont {D.~J.~H.}\
  \bibnamefont {Chung}}, \bibinfo {author} {\bibfnamefont {E.~W.}\ \bibnamefont
  {Kolb}}, \ and\ \bibinfo {author} {\bibfnamefont {A.}~\bibnamefont
  {Riotto}},\ }\href {\doibase 10.1103/PhysRevD.59.023501} {\bibfield
  {journal} {\bibinfo  {journal} {Phys. Rev.}\ }\textbf {\bibinfo {volume}
  {D59}},\ \bibinfo {pages} {023501} (\bibinfo {year} {1999})},\ \Eprint
  {http://arxiv.org/abs/hep-ph/9802238} {arXiv:hep-ph/9802238 [hep-ph]}
  \BibitemShut {NoStop}%
\bibitem [{\citenamefont {Peebles}\ and\ \citenamefont
  {Vilenkin}(1999)}]{Peebles:1999fz}%
  \BibitemOpen
  \bibfield  {author} {\bibinfo {author} {\bibfnamefont {P.~J.~E.}\
  \bibnamefont {Peebles}}\ and\ \bibinfo {author} {\bibfnamefont
  {A.}~\bibnamefont {Vilenkin}},\ }\href {\doibase 10.1103/PhysRevD.60.103506}
  {\bibfield  {journal} {\bibinfo  {journal} {Phys. Rev.}\ }\textbf {\bibinfo
  {volume} {D60}},\ \bibinfo {pages} {103506} (\bibinfo {year} {1999})},\
  \Eprint {http://arxiv.org/abs/astro-ph/9904396} {arXiv:astro-ph/9904396
  [astro-ph]} \BibitemShut {NoStop}%
\bibitem [{\citenamefont {Enqvist}\ \emph {et~al.}(2014)\citenamefont
  {Enqvist}, \citenamefont {Nurmi}, \citenamefont {Tenkanen},\ and\
  \citenamefont {Tuominen}}]{Enqvist:2014zqa}%
  \BibitemOpen
  \bibfield  {author} {\bibinfo {author} {\bibfnamefont {K.}~\bibnamefont
  {Enqvist}}, \bibinfo {author} {\bibfnamefont {S.}~\bibnamefont {Nurmi}},
  \bibinfo {author} {\bibfnamefont {T.}~\bibnamefont {Tenkanen}}, \ and\
  \bibinfo {author} {\bibfnamefont {K.}~\bibnamefont {Tuominen}},\ }\href
  {\doibase 10.1088/1475-7516/2014/08/035} {\bibfield  {journal} {\bibinfo
  {journal} {JCAP}\ }\textbf {\bibinfo {volume} {1408}},\ \bibinfo {pages}
  {035} (\bibinfo {year} {2014})},\ \Eprint {http://arxiv.org/abs/1407.0659}
  {arXiv:1407.0659 [astro-ph.CO]} \BibitemShut {NoStop}%
\bibitem [{\citenamefont {Graham}\ \emph {et~al.}(2016)\citenamefont {Graham},
  \citenamefont {Mardon},\ and\ \citenamefont {Rajendran}}]{Graham:2015rva}%
  \BibitemOpen
  \bibfield  {author} {\bibinfo {author} {\bibfnamefont {P.~W.}\ \bibnamefont
  {Graham}}, \bibinfo {author} {\bibfnamefont {J.}~\bibnamefont {Mardon}}, \
  and\ \bibinfo {author} {\bibfnamefont {S.}~\bibnamefont {Rajendran}},\ }\href
  {\doibase 10.1103/PhysRevD.93.103520} {\bibfield  {journal} {\bibinfo
  {journal} {Phys. Rev.}\ }\textbf {\bibinfo {volume} {D93}},\ \bibinfo {pages}
  {103520} (\bibinfo {year} {2016})},\ \Eprint
  {http://arxiv.org/abs/1504.02102} {arXiv:1504.02102 [hep-ph]} \BibitemShut
  {NoStop}%
\bibitem [{\citenamefont {Nurmi}\ \emph {et~al.}(2015)\citenamefont {Nurmi},
  \citenamefont {Tenkanen},\ and\ \citenamefont {Tuominen}}]{Nurmi:2015ema}%
  \BibitemOpen
  \bibfield  {author} {\bibinfo {author} {\bibfnamefont {S.}~\bibnamefont
  {Nurmi}}, \bibinfo {author} {\bibfnamefont {T.}~\bibnamefont {Tenkanen}}, \
  and\ \bibinfo {author} {\bibfnamefont {K.}~\bibnamefont {Tuominen}},\ }\href
  {\doibase 10.1088/1475-7516/2015/11/001} {\bibfield  {journal} {\bibinfo
  {journal} {JCAP}\ }\textbf {\bibinfo {volume} {1511}},\ \bibinfo {pages}
  {001} (\bibinfo {year} {2015})},\ \Eprint {http://arxiv.org/abs/1506.04048}
  {arXiv:1506.04048 [astro-ph.CO]} \BibitemShut {NoStop}%
\bibitem [{\citenamefont {Garny}\ \emph {et~al.}(2016)\citenamefont {Garny},
  \citenamefont {Sandora},\ and\ \citenamefont {Sloth}}]{Garny:2015sjg}%
  \BibitemOpen
  \bibfield  {author} {\bibinfo {author} {\bibfnamefont {M.}~\bibnamefont
  {Garny}}, \bibinfo {author} {\bibfnamefont {M.}~\bibnamefont {Sandora}}, \
  and\ \bibinfo {author} {\bibfnamefont {M.~S.}\ \bibnamefont {Sloth}},\ }\href
  {\doibase 10.1103/PhysRevLett.116.101302} {\bibfield  {journal} {\bibinfo
  {journal} {Phys. Rev. Lett.}\ }\textbf {\bibinfo {volume} {116}},\ \bibinfo
  {pages} {101302} (\bibinfo {year} {2016})},\ \Eprint
  {http://arxiv.org/abs/1511.03278} {arXiv:1511.03278 [hep-ph]} \BibitemShut
  {NoStop}%
\bibitem [{\citenamefont {Markkanen}\ and\ \citenamefont
  {Nurmi}(2017)}]{Markkanen:2015xuw}%
  \BibitemOpen
  \bibfield  {author} {\bibinfo {author} {\bibfnamefont {T.}~\bibnamefont
  {Markkanen}}\ and\ \bibinfo {author} {\bibfnamefont {S.}~\bibnamefont
  {Nurmi}},\ }\href {\doibase 10.1088/1475-7516/2017/02/008} {\bibfield
  {journal} {\bibinfo  {journal} {JCAP}\ }\textbf {\bibinfo {volume} {1702}},\
  \bibinfo {pages} {008} (\bibinfo {year} {2017})},\ \Eprint
  {http://arxiv.org/abs/1512.07288} {arXiv:1512.07288 [astro-ph.CO]}
  \BibitemShut {NoStop}%
\bibitem [{\citenamefont {Kainulainen}\ \emph {et~al.}(2016)\citenamefont
  {Kainulainen}, \citenamefont {Nurmi}, \citenamefont {Tenkanen}, \citenamefont
  {Tuominen},\ and\ \citenamefont {Vaskonen}}]{Kainulainen:2016vzv}%
  \BibitemOpen
  \bibfield  {author} {\bibinfo {author} {\bibfnamefont {K.}~\bibnamefont
  {Kainulainen}}, \bibinfo {author} {\bibfnamefont {S.}~\bibnamefont {Nurmi}},
  \bibinfo {author} {\bibfnamefont {T.}~\bibnamefont {Tenkanen}}, \bibinfo
  {author} {\bibfnamefont {K.}~\bibnamefont {Tuominen}}, \ and\ \bibinfo
  {author} {\bibfnamefont {V.}~\bibnamefont {Vaskonen}},\ }\href {\doibase
  10.1088/1475-7516/2016/06/022} {\bibfield  {journal} {\bibinfo  {journal}
  {JCAP}\ }\textbf {\bibinfo {volume} {1606}},\ \bibinfo {pages} {022}
  (\bibinfo {year} {2016})},\ \Eprint {http://arxiv.org/abs/1601.07733}
  {arXiv:1601.07733 [astro-ph.CO]} \BibitemShut {NoStop}%
\bibitem [{\citenamefont {Bertolami}\ \emph {et~al.}(2016)\citenamefont
  {Bertolami}, \citenamefont {Cosme},\ and\ \citenamefont
  {Rosa}}]{Bertolami:2016ywc}%
  \BibitemOpen
  \bibfield  {author} {\bibinfo {author} {\bibfnamefont {O.}~\bibnamefont
  {Bertolami}}, \bibinfo {author} {\bibfnamefont {C.}~\bibnamefont {Cosme}}, \
  and\ \bibinfo {author} {\bibfnamefont {J.~G.}\ \bibnamefont {Rosa}},\ }\href
  {\doibase 10.1016/j.physletb.2016.05.047} {\bibfield  {journal} {\bibinfo
  {journal} {Phys. Lett.}\ }\textbf {\bibinfo {volume} {B759}},\ \bibinfo
  {pages} {1} (\bibinfo {year} {2016})},\ \Eprint
  {http://arxiv.org/abs/1603.06242} {arXiv:1603.06242 [hep-ph]} \BibitemShut
  {NoStop}%
\bibitem [{\citenamefont {Heikinheimo}\ \emph {et~al.}(2016)\citenamefont
  {Heikinheimo}, \citenamefont {Tenkanen}, \citenamefont {Tuominen},\ and\
  \citenamefont {Vaskonen}}]{Heikinheimo:2016yds}%
  \BibitemOpen
  \bibfield  {author} {\bibinfo {author} {\bibfnamefont {M.}~\bibnamefont
  {Heikinheimo}}, \bibinfo {author} {\bibfnamefont {T.}~\bibnamefont
  {Tenkanen}}, \bibinfo {author} {\bibfnamefont {K.}~\bibnamefont {Tuominen}},
  \ and\ \bibinfo {author} {\bibfnamefont {V.}~\bibnamefont {Vaskonen}},\
  }\href {\doibase 10.1103/PhysRevD.96.109902, 10.1103/PhysRevD.94.063506}
  {\bibfield  {journal} {\bibinfo  {journal} {Phys. Rev.}\ }\textbf {\bibinfo
  {volume} {D94}},\ \bibinfo {pages} {063506} (\bibinfo {year} {2016})},\
  \bibinfo {note} {[Erratum: Phys. Rev.D96,no.10,109902(2017)]},\ \Eprint
  {http://arxiv.org/abs/1604.02401} {arXiv:1604.02401 [astro-ph.CO]}
  \BibitemShut {NoStop}%
\bibitem [{\citenamefont {Cosme}\ \emph
  {et~al.}(2018{\natexlab{a}})\citenamefont {Cosme}, \citenamefont {Rosa},\
  and\ \citenamefont {Bertolami}}]{Cosme:2017cxk}%
  \BibitemOpen
  \bibfield  {author} {\bibinfo {author} {\bibfnamefont {C.}~\bibnamefont
  {Cosme}}, \bibinfo {author} {\bibfnamefont {J.~G.}\ \bibnamefont {Rosa}}, \
  and\ \bibinfo {author} {\bibfnamefont {O.}~\bibnamefont {Bertolami}},\ }\href
  {\doibase 10.1016/j.physletb.2018.04.062} {\bibfield  {journal} {\bibinfo
  {journal} {Phys. Lett.}\ }\textbf {\bibinfo {volume} {B781}},\ \bibinfo
  {pages} {639} (\bibinfo {year} {2018}{\natexlab{a}})},\ \Eprint
  {http://arxiv.org/abs/1709.09674} {arXiv:1709.09674 [hep-ph]} \BibitemShut
  {NoStop}%
\bibitem [{\citenamefont {Enqvist}\ \emph {et~al.}(2018)\citenamefont
  {Enqvist}, \citenamefont {Hardwick}, \citenamefont {Tenkanen}, \citenamefont
  {Vennin},\ and\ \citenamefont {Wands}}]{Enqvist:2017kzh}%
  \BibitemOpen
  \bibfield  {author} {\bibinfo {author} {\bibfnamefont {K.}~\bibnamefont
  {Enqvist}}, \bibinfo {author} {\bibfnamefont {R.~J.}\ \bibnamefont
  {Hardwick}}, \bibinfo {author} {\bibfnamefont {T.}~\bibnamefont {Tenkanen}},
  \bibinfo {author} {\bibfnamefont {V.}~\bibnamefont {Vennin}}, \ and\ \bibinfo
  {author} {\bibfnamefont {D.}~\bibnamefont {Wands}},\ }\href {\doibase
  10.1088/1475-7516/2018/02/006} {\bibfield  {journal} {\bibinfo  {journal}
  {JCAP}\ }\textbf {\bibinfo {volume} {1802}},\ \bibinfo {pages} {006}
  (\bibinfo {year} {2018})},\ \Eprint {http://arxiv.org/abs/1711.07344}
  {arXiv:1711.07344 [astro-ph.CO]} \BibitemShut {NoStop}%
\bibitem [{\citenamefont {Cosme}\ \emph
  {et~al.}(2018{\natexlab{b}})\citenamefont {Cosme}, \citenamefont {Rosa},\
  and\ \citenamefont {Bertolami}}]{Cosme:2018nly}%
  \BibitemOpen
  \bibfield  {author} {\bibinfo {author} {\bibfnamefont {C.}~\bibnamefont
  {Cosme}}, \bibinfo {author} {\bibfnamefont {J.~G.}\ \bibnamefont {Rosa}}, \
  and\ \bibinfo {author} {\bibfnamefont {O.}~\bibnamefont {Bertolami}},\ }\href
  {\doibase 10.1007/JHEP05(2018)129} {\bibfield  {journal} {\bibinfo  {journal}
  {JHEP}\ }\textbf {\bibinfo {volume} {05}},\ \bibinfo {pages} {129} (\bibinfo
  {year} {2018}{\natexlab{b}})},\ \Eprint {http://arxiv.org/abs/1802.09434}
  {arXiv:1802.09434 [hep-ph]} \BibitemShut {NoStop}%
\bibitem [{\citenamefont {Graham}\ and\ \citenamefont
  {Scherlis}(2018)}]{Graham:2018jyp}%
  \BibitemOpen
  \bibfield  {author} {\bibinfo {author} {\bibfnamefont {P.~W.}\ \bibnamefont
  {Graham}}\ and\ \bibinfo {author} {\bibfnamefont {A.}~\bibnamefont
  {Scherlis}},\ }\href {\doibase 10.1103/PhysRevD.98.035017} {\bibfield
  {journal} {\bibinfo  {journal} {Phys. Rev.}\ }\textbf {\bibinfo {volume}
  {D98}},\ \bibinfo {pages} {035017} (\bibinfo {year} {2018})},\ \Eprint
  {http://arxiv.org/abs/1805.07362} {arXiv:1805.07362 [hep-ph]} \BibitemShut
  {NoStop}%
\bibitem [{\citenamefont {Alonso-Álvarez}\ and\ \citenamefont
  {Jaeckel}(2018)}]{Alonso-Alvarez:2018tus}%
  \BibitemOpen
  \bibfield  {author} {\bibinfo {author} {\bibfnamefont {G.}~\bibnamefont
  {Alonso-Álvarez}}\ and\ \bibinfo {author} {\bibfnamefont {J.}~\bibnamefont
  {Jaeckel}},\ }\href {\doibase 10.1088/1475-7516/2018/10/022} {\bibfield
  {journal} {\bibinfo  {journal} {JCAP}\ }\textbf {\bibinfo {volume} {1810}},\
  \bibinfo {pages} {022} (\bibinfo {year} {2018})},\ \Eprint
  {http://arxiv.org/abs/1807.09785} {arXiv:1807.09785 [hep-ph]} \BibitemShut
  {NoStop}%
\bibitem [{\citenamefont {Ema}\ \emph {et~al.}(2018)\citenamefont {Ema},
  \citenamefont {Nakayama},\ and\ \citenamefont {Tang}}]{Ema:2018ucl}%
  \BibitemOpen
  \bibfield  {author} {\bibinfo {author} {\bibfnamefont {Y.}~\bibnamefont
  {Ema}}, \bibinfo {author} {\bibfnamefont {K.}~\bibnamefont {Nakayama}}, \
  and\ \bibinfo {author} {\bibfnamefont {Y.}~\bibnamefont {Tang}},\ }\href
  {\doibase 10.1007/JHEP09(2018)135} {\bibfield  {journal} {\bibinfo  {journal}
  {JHEP}\ }\textbf {\bibinfo {volume} {09}},\ \bibinfo {pages} {135} (\bibinfo
  {year} {2018})},\ \Eprint {http://arxiv.org/abs/1804.07471} {arXiv:1804.07471
  [hep-ph]} \BibitemShut {NoStop}%
\bibitem [{\citenamefont {Fairbairn}\ \emph {et~al.}(2018)\citenamefont
  {Fairbairn}, \citenamefont {Kainulainen}, \citenamefont {Markkanen},\ and\
  \citenamefont {Nurmi}}]{Fairbairn:2018bsw}%
  \BibitemOpen
  \bibfield  {author} {\bibinfo {author} {\bibfnamefont {M.}~\bibnamefont
  {Fairbairn}}, \bibinfo {author} {\bibfnamefont {K.}~\bibnamefont
  {Kainulainen}}, \bibinfo {author} {\bibfnamefont {T.}~\bibnamefont
  {Markkanen}}, \ and\ \bibinfo {author} {\bibfnamefont {S.}~\bibnamefont
  {Nurmi}},\ }\href@noop {} {\  (\bibinfo {year} {2018})},\ \Eprint
  {http://arxiv.org/abs/1808.08236} {arXiv:1808.08236 [astro-ph.CO]}
  \BibitemShut {NoStop}%
\bibitem [{\citenamefont {Akrami}\ \emph {et~al.}(2018)\citenamefont {Akrami}
  \emph {et~al.}}]{Akrami:2018odb}%
  \BibitemOpen
  \bibfield  {author} {\bibinfo {author} {\bibfnamefont {Y.}~\bibnamefont
  {Akrami}} \emph {et~al.} (\bibinfo {collaboration} {Planck}),\ }\href@noop {}
  {\  (\bibinfo {year} {2018})},\ \Eprint {http://arxiv.org/abs/1807.06211}
  {arXiv:1807.06211 [astro-ph.CO]} \BibitemShut {NoStop}%
\bibitem [{\citenamefont {Marsh}(2016)}]{Marsh:2015xka}%
  \BibitemOpen
  \bibfield  {author} {\bibinfo {author} {\bibfnamefont {D.~J.~E.}\
  \bibnamefont {Marsh}},\ }\href {\doibase 10.1016/j.physrep.2016.06.005}
  {\bibfield  {journal} {\bibinfo  {journal} {Phys. Rept.}\ }\textbf {\bibinfo
  {volume} {643}},\ \bibinfo {pages} {1} (\bibinfo {year} {2016})},\ \Eprint
  {http://arxiv.org/abs/1510.07633} {arXiv:1510.07633 [astro-ph.CO]}
  \BibitemShut {NoStop}%
\bibitem [{\citenamefont {Takahashi}\ \emph {et~al.}(2018)\citenamefont
  {Takahashi}, \citenamefont {Yin},\ and\ \citenamefont {Guth}}]{Guth:2018hsa}%
  \BibitemOpen
  \bibfield  {author} {\bibinfo {author} {\bibfnamefont {F.}~\bibnamefont
  {Takahashi}}, \bibinfo {author} {\bibfnamefont {W.}~\bibnamefont {Yin}}, \
  and\ \bibinfo {author} {\bibfnamefont {A.~H.}\ \bibnamefont {Guth}},\ }\href
  {\doibase 10.1103/PhysRevD.98.015042} {\bibfield  {journal} {\bibinfo
  {journal} {Phys. Rev.}\ }\textbf {\bibinfo {volume} {D98}},\ \bibinfo {pages}
  {015042} (\bibinfo {year} {2018})},\ \Eprint
  {http://arxiv.org/abs/1805.08763} {arXiv:1805.08763 [hep-ph]} \BibitemShut
  {NoStop}%
\bibitem [{\citenamefont {Starobinsky}\ and\ \citenamefont
  {Yokoyama}(1994)}]{Starobinsky:1994bd}%
  \BibitemOpen
  \bibfield  {author} {\bibinfo {author} {\bibfnamefont {A.~A.}\ \bibnamefont
  {Starobinsky}}\ and\ \bibinfo {author} {\bibfnamefont {J.}~\bibnamefont
  {Yokoyama}},\ }\href {\doibase 10.1103/PhysRevD.50.6357} {\bibfield
  {journal} {\bibinfo  {journal} {Phys. Rev.}\ }\textbf {\bibinfo {volume}
  {D50}},\ \bibinfo {pages} {6357} (\bibinfo {year} {1994})},\ \Eprint
  {http://arxiv.org/abs/astro-ph/9407016} {arXiv:astro-ph/9407016 [astro-ph]}
  \BibitemShut {NoStop}%
\bibitem [{\citenamefont {Enqvist}\ \emph {et~al.}(2012)\citenamefont
  {Enqvist}, \citenamefont {Lerner}, \citenamefont {Taanila},\ and\
  \citenamefont {Tranberg}}]{Enqvist:2012xn}%
  \BibitemOpen
  \bibfield  {author} {\bibinfo {author} {\bibfnamefont {K.}~\bibnamefont
  {Enqvist}}, \bibinfo {author} {\bibfnamefont {R.~N.}\ \bibnamefont {Lerner}},
  \bibinfo {author} {\bibfnamefont {O.}~\bibnamefont {Taanila}}, \ and\
  \bibinfo {author} {\bibfnamefont {A.}~\bibnamefont {Tranberg}},\ }\href
  {\doibase 10.1088/1475-7516/2012/10/052} {\bibfield  {journal} {\bibinfo
  {journal} {JCAP}\ }\textbf {\bibinfo {volume} {1210}},\ \bibinfo {pages}
  {052} (\bibinfo {year} {2012})},\ \Eprint {http://arxiv.org/abs/1205.5446}
  {arXiv:1205.5446 [astro-ph.CO]} \BibitemShut {NoStop}%
\bibitem [{\citenamefont {Kunimitsu}\ and\ \citenamefont
  {Yokoyama}(2012)}]{Kunimitsu:2012xx}%
  \BibitemOpen
  \bibfield  {author} {\bibinfo {author} {\bibfnamefont {T.}~\bibnamefont
  {Kunimitsu}}\ and\ \bibinfo {author} {\bibfnamefont {J.}~\bibnamefont
  {Yokoyama}},\ }\href {\doibase 10.1103/PhysRevD.86.083541} {\bibfield
  {journal} {\bibinfo  {journal} {Phys. Rev.}\ }\textbf {\bibinfo {volume}
  {D86}},\ \bibinfo {pages} {083541} (\bibinfo {year} {2012})},\ \Eprint
  {http://arxiv.org/abs/1208.2316} {arXiv:1208.2316 [hep-ph]} \BibitemShut
  {NoStop}%
\bibitem [{\citenamefont {Hardwick}\ \emph {et~al.}(2017)\citenamefont
  {Hardwick}, \citenamefont {Vennin}, \citenamefont {Byrnes}, \citenamefont
  {Torrado},\ and\ \citenamefont {Wands}}]{Hardwick:2017fjo}%
  \BibitemOpen
  \bibfield  {author} {\bibinfo {author} {\bibfnamefont {R.~J.}\ \bibnamefont
  {Hardwick}}, \bibinfo {author} {\bibfnamefont {V.}~\bibnamefont {Vennin}},
  \bibinfo {author} {\bibfnamefont {C.~T.}\ \bibnamefont {Byrnes}}, \bibinfo
  {author} {\bibfnamefont {J.}~\bibnamefont {Torrado}}, \ and\ \bibinfo
  {author} {\bibfnamefont {D.}~\bibnamefont {Wands}},\ }\href {\doibase
  10.1088/1475-7516/2017/10/018} {\bibfield  {journal} {\bibinfo  {journal}
  {JCAP}\ }\textbf {\bibinfo {volume} {1710}},\ \bibinfo {pages} {018}
  (\bibinfo {year} {2017})},\ \Eprint {http://arxiv.org/abs/1701.06473}
  {arXiv:1701.06473 [astro-ph.CO]} \BibitemShut {NoStop}%
\bibitem [{\citenamefont {Ichikawa}\ \emph {et~al.}(2008)\citenamefont
  {Ichikawa}, \citenamefont {Suyama}, \citenamefont {Takahashi},\ and\
  \citenamefont {Yamaguchi}}]{Ichikawa:2008ne}%
  \BibitemOpen
  \bibfield  {author} {\bibinfo {author} {\bibfnamefont {K.}~\bibnamefont
  {Ichikawa}}, \bibinfo {author} {\bibfnamefont {T.}~\bibnamefont {Suyama}},
  \bibinfo {author} {\bibfnamefont {T.}~\bibnamefont {Takahashi}}, \ and\
  \bibinfo {author} {\bibfnamefont {M.}~\bibnamefont {Yamaguchi}},\ }\href
  {\doibase 10.1103/PhysRevD.78.063545} {\bibfield  {journal} {\bibinfo
  {journal} {Phys. Rev.}\ }\textbf {\bibinfo {volume} {D78}},\ \bibinfo {pages}
  {063545} (\bibinfo {year} {2008})},\ \Eprint {http://arxiv.org/abs/0807.3988}
  {arXiv:0807.3988 [astro-ph]} \BibitemShut {NoStop}%
\bibitem [{\citenamefont {Carlson}\ \emph {et~al.}(1992)\citenamefont
  {Carlson}, \citenamefont {Machacek},\ and\ \citenamefont
  {Hall}}]{Carlson:1992fn}%
  \BibitemOpen
  \bibfield  {author} {\bibinfo {author} {\bibfnamefont {E.~D.}\ \bibnamefont
  {Carlson}}, \bibinfo {author} {\bibfnamefont {M.~E.}\ \bibnamefont
  {Machacek}}, \ and\ \bibinfo {author} {\bibfnamefont {L.~J.}\ \bibnamefont
  {Hall}},\ }\href {\doibase 10.1086/171833} {\bibfield  {journal} {\bibinfo
  {journal} {Astrophys. J.}\ }\textbf {\bibinfo {volume} {398}},\ \bibinfo
  {pages} {43} (\bibinfo {year} {1992})}\BibitemShut {NoStop}%
\bibitem [{\citenamefont {Tenkanen}\ and\ \citenamefont
  {Vaskonen}(2016)}]{Tenkanen:2016jic}%
  \BibitemOpen
  \bibfield  {author} {\bibinfo {author} {\bibfnamefont {T.}~\bibnamefont
  {Tenkanen}}\ and\ \bibinfo {author} {\bibfnamefont {V.}~\bibnamefont
  {Vaskonen}},\ }\href {\doibase 10.1103/PhysRevD.94.083516} {\bibfield
  {journal} {\bibinfo  {journal} {Phys. Rev.}\ }\textbf {\bibinfo {volume}
  {D94}},\ \bibinfo {pages} {083516} (\bibinfo {year} {2016})},\ \Eprint
  {http://arxiv.org/abs/1606.00192} {arXiv:1606.00192 [astro-ph.CO]}
  \BibitemShut {NoStop}%
\bibitem [{\citenamefont {Ade}\ \emph {et~al.}(2018)\citenamefont {Ade} \emph
  {et~al.}}]{Ade:2018gkx}%
  \BibitemOpen
  \bibfield  {author} {\bibinfo {author} {\bibfnamefont {P.~A.~R.}\
  \bibnamefont {Ade}} \emph {et~al.} (\bibinfo {collaboration} {BICEP2, Keck
  Array}),\ }\href@noop {} {\bibfield  {journal} {\bibinfo  {journal}
  {Submitted to: Phys. Rev. Lett.}\ } (\bibinfo {year} {2018})},\ \Eprint
  {http://arxiv.org/abs/1810.05216} {arXiv:1810.05216 [astro-ph.CO]}
  \BibitemShut {NoStop}%
\bibitem [{\citenamefont {Matsumura}\ \emph {et~al.}(2013)\citenamefont
  {Matsumura} \emph {et~al.}}]{Matsumura:2013aja}%
  \BibitemOpen
  \bibfield  {author} {\bibinfo {author} {\bibfnamefont {T.}~\bibnamefont
  {Matsumura}} \emph {et~al.},\ }\href {\doibase 10.1007/s10909-013-0996-1} {\
  (\bibinfo {year} {2013}),\ 10.1007/s10909-013-0996-1},\ \bibinfo {note} {[J.
  Low. Temp. Phys.176,733(2014)]},\ \Eprint {http://arxiv.org/abs/1311.2847}
  {arXiv:1311.2847 [astro-ph.IM]} \BibitemShut {NoStop}%
\bibitem [{\citenamefont {Wu}\ \emph {et~al.}(2016)\citenamefont {Wu} \emph
  {et~al.}}]{Wu:2016hul}%
  \BibitemOpen
  \bibfield  {author} {\bibinfo {author} {\bibfnamefont {W.~L.~K.}\
  \bibnamefont {Wu}} \emph {et~al.},\ }\href {\doibase
  10.1007/s10909-015-1403-x} {\bibfield  {journal} {\bibinfo  {journal} {J.
  Low. Temp. Phys.}\ }\textbf {\bibinfo {volume} {184}},\ \bibinfo {pages}
  {765} (\bibinfo {year} {2016})},\ \Eprint {http://arxiv.org/abs/1601.00125}
  {arXiv:1601.00125 [astro-ph.IM]} \BibitemShut {NoStop}%
\bibitem [{\citenamefont {Abazajian}\ \emph {et~al.}(2016)\citenamefont
  {Abazajian} \emph {et~al.}}]{Abazajian:2016yjj}%
  \BibitemOpen
  \bibfield  {author} {\bibinfo {author} {\bibfnamefont {K.~N.}\ \bibnamefont
  {Abazajian}} \emph {et~al.} (\bibinfo {collaboration} {CMB-S4}),\ }\href@noop
  {} {\  (\bibinfo {year} {2016})},\ \Eprint {http://arxiv.org/abs/1610.02743}
  {arXiv:1610.02743 [astro-ph.CO]} \BibitemShut {NoStop}%
\bibitem [{\citenamefont {Aguirre}\ \emph {et~al.}(2018)\citenamefont {Aguirre}
  \emph {et~al.}}]{Ade:2018sbj}%
  \BibitemOpen
  \bibfield  {author} {\bibinfo {author} {\bibfnamefont {J.}~\bibnamefont
  {Aguirre}} \emph {et~al.} (\bibinfo {collaboration} {Simons Observatory}),\
  }\href@noop {} {\  (\bibinfo {year} {2018})},\ \Eprint
  {http://arxiv.org/abs/1808.07445} {arXiv:1808.07445 [astro-ph.CO]}
  \BibitemShut {NoStop}%
\bibitem [{\citenamefont {Markevitch}\ \emph {et~al.}(2004)\citenamefont
  {Markevitch}, \citenamefont {Gonzalez}, \citenamefont {Clowe}, \citenamefont
  {Vikhlinin}, \citenamefont {David}, \citenamefont {Forman}, \citenamefont
  {Jones}, \citenamefont {Murray},\ and\ \citenamefont
  {Tucker}}]{Markevitch:2003at}%
  \BibitemOpen
  \bibfield  {author} {\bibinfo {author} {\bibfnamefont {M.}~\bibnamefont
  {Markevitch}}, \bibinfo {author} {\bibfnamefont {A.~H.}\ \bibnamefont
  {Gonzalez}}, \bibinfo {author} {\bibfnamefont {D.}~\bibnamefont {Clowe}},
  \bibinfo {author} {\bibfnamefont {A.}~\bibnamefont {Vikhlinin}}, \bibinfo
  {author} {\bibfnamefont {L.}~\bibnamefont {David}}, \bibinfo {author}
  {\bibfnamefont {W.}~\bibnamefont {Forman}}, \bibinfo {author} {\bibfnamefont
  {C.}~\bibnamefont {Jones}}, \bibinfo {author} {\bibfnamefont
  {S.}~\bibnamefont {Murray}}, \ and\ \bibinfo {author} {\bibfnamefont
  {W.}~\bibnamefont {Tucker}},\ }\href {\doibase 10.1086/383178} {\bibfield
  {journal} {\bibinfo  {journal} {Astrophys. J.}\ }\textbf {\bibinfo {volume}
  {606}},\ \bibinfo {pages} {819} (\bibinfo {year} {2004})},\ \Eprint
  {http://arxiv.org/abs/astro-ph/0309303} {arXiv:astro-ph/0309303 [astro-ph]}
  \BibitemShut {NoStop}%
\bibitem [{\citenamefont {Randall}\ \emph {et~al.}(2008)\citenamefont
  {Randall}, \citenamefont {Markevitch}, \citenamefont {Clowe}, \citenamefont
  {Gonzalez},\ and\ \citenamefont {Bradac}}]{Randall:2007ph}%
  \BibitemOpen
  \bibfield  {author} {\bibinfo {author} {\bibfnamefont {S.~W.}\ \bibnamefont
  {Randall}}, \bibinfo {author} {\bibfnamefont {M.}~\bibnamefont {Markevitch}},
  \bibinfo {author} {\bibfnamefont {D.}~\bibnamefont {Clowe}}, \bibinfo
  {author} {\bibfnamefont {A.~H.}\ \bibnamefont {Gonzalez}}, \ and\ \bibinfo
  {author} {\bibfnamefont {M.}~\bibnamefont {Bradac}},\ }\href {\doibase
  10.1086/587859} {\bibfield  {journal} {\bibinfo  {journal} {Astrophys. J.}\
  }\textbf {\bibinfo {volume} {679}},\ \bibinfo {pages} {1173} (\bibinfo {year}
  {2008})},\ \Eprint {http://arxiv.org/abs/0704.0261} {arXiv:0704.0261
  [astro-ph]} \BibitemShut {NoStop}%
\bibitem [{\citenamefont {Rocha}\ \emph {et~al.}(2013)\citenamefont {Rocha},
  \citenamefont {Peter}, \citenamefont {Bullock}, \citenamefont {Kaplinghat},
  \citenamefont {Garrison-Kimmel}, \citenamefont {Onorbe},\ and\ \citenamefont
  {Moustakas}}]{Rocha:2012jg}%
  \BibitemOpen
  \bibfield  {author} {\bibinfo {author} {\bibfnamefont {M.}~\bibnamefont
  {Rocha}}, \bibinfo {author} {\bibfnamefont {A.~H.~G.}\ \bibnamefont {Peter}},
  \bibinfo {author} {\bibfnamefont {J.~S.}\ \bibnamefont {Bullock}}, \bibinfo
  {author} {\bibfnamefont {M.}~\bibnamefont {Kaplinghat}}, \bibinfo {author}
  {\bibfnamefont {S.}~\bibnamefont {Garrison-Kimmel}}, \bibinfo {author}
  {\bibfnamefont {J.}~\bibnamefont {Onorbe}}, \ and\ \bibinfo {author}
  {\bibfnamefont {L.~A.}\ \bibnamefont {Moustakas}},\ }\href {\doibase
  10.1093/mnras/sts514} {\bibfield  {journal} {\bibinfo  {journal} {Mon. Not.
  Roy. Astron. Soc.}\ }\textbf {\bibinfo {volume} {430}},\ \bibinfo {pages}
  {81} (\bibinfo {year} {2013})},\ \Eprint {http://arxiv.org/abs/1208.3025}
  {arXiv:1208.3025 [astro-ph.CO]} \BibitemShut {NoStop}%
\bibitem [{\citenamefont {Peter}\ \emph {et~al.}(2013)\citenamefont {Peter},
  \citenamefont {Rocha}, \citenamefont {Bullock},\ and\ \citenamefont
  {Kaplinghat}}]{Peter:2012jh}%
  \BibitemOpen
  \bibfield  {author} {\bibinfo {author} {\bibfnamefont {A.~H.~G.}\
  \bibnamefont {Peter}}, \bibinfo {author} {\bibfnamefont {M.}~\bibnamefont
  {Rocha}}, \bibinfo {author} {\bibfnamefont {J.~S.}\ \bibnamefont {Bullock}},
  \ and\ \bibinfo {author} {\bibfnamefont {M.}~\bibnamefont {Kaplinghat}},\
  }\href {\doibase 10.1093/mnras/sts535} {\bibfield  {journal} {\bibinfo
  {journal} {Mon. Not. Roy. Astron. Soc.}\ }\textbf {\bibinfo {volume} {430}},\
  \bibinfo {pages} {105} (\bibinfo {year} {2013})},\ \Eprint
  {http://arxiv.org/abs/1208.3026} {arXiv:1208.3026 [astro-ph.CO]} \BibitemShut
  {NoStop}%
\bibitem [{\citenamefont {Harvey}\ \emph {et~al.}(2015)\citenamefont {Harvey},
  \citenamefont {Massey}, \citenamefont {Kitching}, \citenamefont {Taylor},\
  and\ \citenamefont {Tittley}}]{Harvey:2015hha}%
  \BibitemOpen
  \bibfield  {author} {\bibinfo {author} {\bibfnamefont {D.}~\bibnamefont
  {Harvey}}, \bibinfo {author} {\bibfnamefont {R.}~\bibnamefont {Massey}},
  \bibinfo {author} {\bibfnamefont {T.}~\bibnamefont {Kitching}}, \bibinfo
  {author} {\bibfnamefont {A.}~\bibnamefont {Taylor}}, \ and\ \bibinfo {author}
  {\bibfnamefont {E.}~\bibnamefont {Tittley}},\ }\href {\doibase
  10.1126/science.1261381} {\bibfield  {journal} {\bibinfo  {journal}
  {Science}\ }\textbf {\bibinfo {volume} {347}},\ \bibinfo {pages} {1462}
  (\bibinfo {year} {2015})},\ \Eprint {http://arxiv.org/abs/1503.07675}
  {arXiv:1503.07675 [astro-ph.CO]} \BibitemShut {NoStop}%
\end{thebibliography}%

\end{document}